\documentclass[aip,  amsmath,amssymb, reprint, floatfix]{revtex4-1}
\usepackage{graphicx}
\usepackage{dcolumn}
\usepackage{float}
\usepackage{bm}
\usepackage[dvipsnames,table,xcdraw]{xcolor}
\usepackage[utf8]{inputenc}
\usepackage[T1]{fontenc}
\usepackage{mathptmx}
\usepackage{mathtools}

\begin{document}
\title{Effects of a Velocity Shear on Double Current Sheet Systems: Explosive Reconnection and Particle Acceleration}

\author{Arghyadeep Paul$^{*}$}
\email[]{arghyadeepp@gmail.com ($^{*}$Corresponding Author)}
\affiliation{Department of Astronomy Astrophysics and Space Engineering, Indian Institute of Technology Indore, Khandwa Road, Simrol, Indore 453552, India}

\author{Bhargav Vaidya}
\email[]{bvaidya@iiti.ac.in}
\affiliation{Department of Astronomy Astrophysics and Space Engineering, Indian Institute of Technology Indore, Khandwa Road, Simrol, Indore 453552, India}


\begin{abstract}The effect of a parallel velocity shear on the explosive phase of a double current sheet system is investigated within the 2D resistive magnetohydrodynamic (MHD) framework. We further explore the effect of this shear on acceleration of test particles. The general evolution pattern of the double current sheets is similar for all sub-Alfv\'enic shears with respect to the initial transient phase, the onset of the plasmoid instability and the final relaxation phase. We find that the theoretical scaling of the reconnection rate with shear holds if the rate is measured when the islands have a similar size. The larger island widths for lower shears greatly enhance the reconnection rate during the explosive phase. We have further examined the modification of the energy spectrum of the accelerated particles in the presence of a shear. Our results also show that the flow only modifies the high energy tail of the particle spectrum and has negligible effect on the power-law index. Individual particle trajectories help to explore the various mechanisms associated with the acceleration. Based on the location of the particles, the acceleration mechanisms are found to vary. We highlight the importance of the convective electric field in the inflow as well as the outflow region inside large magnetic islands in the acceleration of particles. The interaction and reflection of the particles with the reconnection exhausts inside the large scale primary magnetic islands is found to have a significant effect on the energization of the particles.    
\end{abstract}

\maketitle 

\section{\label{Intro}Introduction:\protect}

Magnetic reconnection is a ubiquitous phenomenon in astrophysical and laboratory plasmas where a rearrangement of topology of the magnetic field lines leads to an overall magnetic relaxation of the system. This rearrangement occurs at a very localized region of the plasma domain where non-ideal effects dominate and break the `frozen-in' condition of the magnetic field lines. The role of magnetic reconnection in the astrophysical scenario leading to the conversion of magnetic energy into thermal and kinetic energy was initially motivated by the study of solar flares\cite{Giovanelli_1939, Giovanelli_1947, Giovanelli_1948, Hesse_2020}. Reconnection is also believed to play a pivotal role in the ejection of Coronal Mass Ejections (CMEs) from the solar surface\cite{Karpen_2012}. These CMEs are ejected into the ambient solar wind where the interplay of the magnetic fields of the CMEs with that of the solar wind may also lead to the formation of potential reconnection regions as suggested by observations and simulations\cite{Gosling_1995, Schmidt_2003}. This process of magnetic-reconnection also has direct implications at Earth as the same phenomena responsible for such energetic events at the solar surface is also responsible for the Sun-Earth connection causing the inflow of the solar wind plasma and Solar Energetic Particles (SEPs) into the Earth's magnetospheric system\cite{Dungey_1961}. Some enduring issues in the context of magnetic reconnection are the partitioning of magnetic energy during the energy conversion process\cite{Fadanelli_2021} and the production of high energy particles by non-thermal acceleration mechanisms close to the reconnection sites\cite{hossienpour_plasmoid_shear, Drake_2006}. The observations and measurement of fast exhaust outflows and high energy particles from the reconnection regions in various astrophysical and laboratory environments such as solar flares, magnetospheric substorms, tokamaks etc. provide direct evidence of this energy conversion process\cite{Mann_2009, Hesse_2020, Romanova_1992, dal_Pino_2005}.

The well known Sweet-Parker model of magnetic reconnection predicts the reconnection rates to scale as $S^{-1/2}$ where $S=Lv_{\rm A}/\eta$ is the Lundquist number of the system having a length scale of L and an Alfv\'en speed of $v_{\rm A}$ with $\eta$ being the resistivity. The reconnection time-scales estimated by the Sweet-Parker model\cite{sweet_parker1, sweet_parker_2} is however, inconsistent with the observations available in the heliophysics domain as the high Lundquist number of such systems (typically as high as $10^{10-14}$) render the theoretical time-scale to be too slow. The Petschek model improves on this with a weaker dependence on S where the maximum reconnection rate is found to scale as $\sim 1/\log S$. But studies have shown that the Petschek type reconnection is only achievable with an additional requirement of a localized enhancement of resistivity near the current sheets\cite{biskamp_2000, huang_bhattacharjee_2010}. Studies on Petschek type reconnection with uniform resistivity have revealed the marginal stability of such systems\cite{Baty_2009}. It has recently been realized that the emergence of a secondary tearing instability, called the `plasmoid instability', can occur in high Lundquist number (S>$10^4$) Sweet-Parker current sheets which can greatly enhance the reconnection rate by fragmenting the current sheet into multiple X-points\cite{Biskamp_1986, Daughton_2006, Drake_2006,Loureiro_2007, Lapenta_2008, Lin_2008, Bhattacharjee_2009, Nemati_2015}. It has also been found that in this regime, the reconnection rate is practically independent of the Lundquist number\cite{huang_bhattacharjee_2010}. Plasmoid instability has been hypothesized to be an important factor from the perspective of fast magnetic reconnection and is thus of paramount importance in the context of sawtooth crashes in laboratory plasmas, reconnection in the solar corona, and flux transfer events at the day-side magneto-pause\cite{Yu_2014, Lee_1986, Lee_Fu_1985}. Multiple observations of various reconnection environments in the heliosphere have also extensively confirmed the presence of such plasmoids\cite{Antolin_2020, Li_2016, Gunell_2012, Lemaire_1981}. 

A detailed study of various reconnection environments dictates that the presence of multiple current sheets is quite prevalent and has been realized in the context of heliospheric plasma\cite{Crooker_1993, Karpen_1996}. The same notion can also been extended to pulsar magnetospheres\cite{Baty_2013} and Active Galactic Nuclei (AGN) jets\cite{Giannios_2009}. Recently, the study of the double tearing mode or DTM has gained significant attention due to its ability to exhibit a structure driven fast non-linear growth phase with a time scale that is weakly dependent on resistivity. This is an important characteristic as the onset of fast magnetic reconnection with a weak dependence on $\eta$ can consequently be emulated in a simple resistive-MHD scenario\cite{Janvier2011}. The plasmoid instability in DTMs is also an important factor in further enhancing the reconnection rate. The formation of small scale plasmoids after the fast flux-driven DTM reconnection phase has been found to drastically increase the kinetic energy of the system and the growth rate of the large plasmoids was shown to increase with an increasing value of the Lundquist number\cite{nemati_2016}. 

The highly time varying condition in the heliospheric reconnection environments also suggests the presence of a velocity shear to be quite common near the current sheets. These shear flows may either be generated externally due to a separate driving force or be spontaneously self-generated within system\cite{Mao_2013, Wang_2008}. Velocity shear flows across the current sheets in the direction parallel to the magnetic field lines usually tend to throttle the growth of the tearing instability which essentially result in a suppression of the reconnection rate\cite{mitchell_kan_1978}. This has important implications in the magnetospheric context as the speed of the solar wind near the Earth's magnetopause can give rise to either sub-Alfv\'enic\cite{Gosling_1982} or super-Alfv\'enic\cite{Belcher_1971} shear speeds across the magnetic reconnection regions at the magnetopause and the same has been observed extensively\cite{Cassak_otto_2011, Gosling_1991, Gosling_1996, Kessel_1996, Onsager_2001, Avanov_2001, Fedorov_2001}. Early theoretical calculations have predicted that reconnection should not occur if the velocity shear is super-Alfv\'enic\cite{mitchell_kan_1978, Labellehamer_1994}. Simulations of magnetic-reconnection for Harris-sheet like configurations in the presence of a shear flow confirmed that steady state reconnection can indeed not be achieved in the presence of a super-Alfv\'enic shear flow\cite{Labellehamer_1994}. Simulations for slower sub-Alfv\'enic flows have also shown that the reconnection rates tend to decrease with an increasing shear\cite{Chen_1997} and a quantitative formulation of the same was given by Cassak and Otto (2011)\cite{Cassak_otto_2011}. It has also been highlighted in literature that a sub-Alfv\'enic shear can either boost or suppress the tearing mode instability and this effect is dependent on the thickness of the shear flow\cite{Li_2010}. Recent studies with particle-in-cell simulations have shown the effects of a combination of an asymmetric magnetic field, a guide field, and the presence of a shear flow in the context of magnetopause reconnection\cite{Tanaka2010}. The effect of a shear flow on the plasmoid instability in a single current sheet has also been lately studied by Hosseinpour et.al.\cite{hossienpour_plasmoid_shear} and it was found that the scaling relation between the shear speed and the reconnection rate given by Cassak and Otto \cite{Cassak_otto_2011} for the tearing instability is also applicable for the plasmoid instability. The plasmoid instability phase in the secondary current sheets of DTMs has an additional complexity caused by the resonant flux feedback of the two current sheets affecting the overall reconnection rate of the system. Whether the scaling of the reconnection rate with shear is valid for such an explosive non-linear phase of the DTM is still unexplored.

The phenomena of particle acceleration to suprathermal energies is also an important signature of energy release in reconnection regions. These kinds of suprathermal particles, consisting of both electrons and ions have been observed in solar flares\cite{Laming_2013}, in the solar wind\cite{Fisk_2006}, AGN jets\cite{Li_2010, Giannios_2009, Striani_2016} and various other regions where reconnection is expected to be a key factor in energy conversion. Interestingly, in most of the cases, the particles have an energy spectrum in the form of a power-law\cite{Drake_2013} that depends on the energy (E) as $E^{-1.5}$. Many acceleration mechanisms such as the stochastic acceleration by MHD waves, direct electric field acceleration by both sub and super-Dreicer electric fields\cite{Miller_1997}, Fermi mechanisms, and acceleration by reconnection driven turbulence\cite{Zharkova2011} has been proposed to explain the production of high-energy particles in these regions. Observational signatures of such accelerated particles include the hard X-ray (HXR) looptop sources that are caused by the trapped high energy particles\cite{Masuda_1994, Somov_1997} and the detection of solar energetic particle (SEP) events at 1 AU as well\cite{Lin_2011}. Recent studies involving observations and simulations have shown that the reconnection in solar flares involves multiple X-points and plasmoids\cite{Drake_2013,Daughton_2011, Fermo_2012} and thus, a comprehensive study of particle acceleration in such regions must take their effects into account. Drake et. al \cite{Drake_2013} have highlighted that the $E^{-1.5}$ spectrum is a universal feature in a multi-island magnetic reconnection system where the acceleration time of the particles are shorter than their loss rate albeit the kinetic equations governing the particle motions were specified in the non-relativistic limit. Test particle simulations in a similar non-relativistic limit by Akramov and Baty (2017) has also previously been used to describe the dynamics of particle acceleration in the explosive phase of DTM reconnection\cite{akramov_baty}, however the particle integrations were done in a static fluid background taken as a snapshot of the domain during the respective phase. This method inherently gives rise to a limitation in the duration of the time evolution of particles in addition to the concern that capturing the effects caused by fast transient structures evolving in the dynamic background fluid on the particles become challenging. Thus, a dynamic fluid background is preferable for such fast evolving systems as the consideration of a static background, may in some cases, lead to an underestimation in the energy growth of the particles. Despite of the extensive research so far, the exact mechanisms governing the particle acceleration in such single or multi current layer reconnection systems\cite{dal_Pino_2005} dominated by fast moving plasmoids is not very well understood.

In this paper, we perform resistive-MHD simulations of very high Lundquist number current sheets in a DTM configuration with the presence of a shear flow across the current layers with an aim to highlight the effects of such a shear flow in the linear and explosive non-linear phases of the DTM like evolution process.The relevance of our work in this paper significantly leans towards the reconnection environments like those seen in the solar corona. We also perform test particle simulations in a dynamic fluid background in an effort to provide insights into the particle acceleration mechanisms during the explosive phases of such systems. The paper is organized as follows: Section \ref{Setup}
describes the model setup and the numerical framework along with the initial conditions; Section \ref{Results} describe the principal findings of the study and Section \ref{Summary} contains the summary along with some additional discussions relevant to the study.

\section{\label{Setup}Model Setup:\protect}

\subsection{\label{Fluid Setup}Numerical framework and Initial conditions:}

The numerical framework consists of solving a standard set of resistive MHD equations in 2-dimensions using the MHD module of the PLUTO code \cite{Mignone_2007}. The MHD module solves a set of conservation laws which are given as:

\begin{equation}
\begin{split}
    \frac{\partial \rho}{\partial t} &+ \nabla .(\rho \textbf{v}) = 0 \\
    \frac{\partial \textbf{m}}{\partial t} &+ \nabla. \left[ \textbf{mv} -\textbf{BB} \right] + \nabla\left(p + \frac{B^2}{2}\right) = 0\\
    \frac{\partial \textbf{B}}{\partial t} &+ \nabla \times \left( c \textbf{E}\right) = 0
\end{split}
\end{equation}

where $\rho$ is the mass density, \textbf{v} is the velocity,  \textbf{m}=$\rho \textbf{v}$ is thus, the momentum density. In the momentum conservation equation, the terms \textbf{mv} and \textbf{BB} represents the corresponding vector outer product. The system is considered to be isothermal and hence, the energy equation is ignored and thermal pressure is derived from the density using the relation $\textit{p}= \rho c_{\rm iso}^2$ where $c_{\rm iso}$ is the isothermal sound speed which was set to be equal to unity in our case. In the third equation, i.e the induction equation, the electric field $\textbf{E}$ is defined by

\begin{equation}\label{elec_F}
    c\textbf{E}= -\textbf{v} \times \textbf{B} + \frac{\eta}{c}\cdot \mathbf{J}
\end{equation}

where $\eta$ is the plasma resistivity which was set to be constant for all our runs having a value of $\eta = 2\times 10^{-5}$ and $\mathbf{J} \equiv c\nabla \times \mathbf{B}$ is the current density. {Also note that a factor of $1/\sqrt{4\pi}$ has been absorbed in the definition of \textbf{B} and \textbf{J}.}

Our initial configuration is that of a double tearing mode (DTM) with the reconnecting magnetic field parallel to the x-axis with a variation along the y-direction. The profile is mathematically given as:

\begin{equation}
    \begin{split}
        B_{x}(y) &= B_0 \left[\tanh\left( \frac{y-l}{w_B}\right) - \tanh\left( \frac{y+l}{w_B}\right) +1 \right]\\
        B_{y} &= 0
    \end{split}
\end{equation}

where $B_0= \sqrt{2}$ is the maximum amplitude of the reconnecting magnetic field, $w_B=1.0$ is the parameter which defines the width over which the magnetic field changes sign and the value of $l= 16.0$ defines the offset of the current sheets from the y=0 line. To achieve an initial equilibrium configuration, the density is set to be a double Harris-sheet profile prescribed as:

\begin{equation}
    \rho(y)= \rho_{\infty}\left[ {\rm sech}^2\left( \frac{y-l}{w_B}\right)+  {\rm sech}^2\left( \frac{y+l}{w_B}\right) +1 \right]
\end{equation}

where the value of $\rho_{\infty}$ is set to be 1.0.
The asymptotic Alfv\'en speed of the system is then defined as $v_{\rm A}= B_0/\sqrt{\rho_{\infty}}= \sqrt{2}$ and the corresponding Lundquist number (S) is given by $S \sim 9\times10^6$. This also gives us a convenient way to normalize our time scale to the Alfv\'en transit time across the current sheet which is defined as $\tau_{\rm A}= w_B/ v_{\rm A}$. The velocity profile was set to be mathematically similar to the magnetic field profile given as 

\begin{equation}
    v_{x}(y) = v_{\rm s} \left[\tanh\left( \frac{y-l}{w_v}\right) - \tanh\left( \frac{y+l}{w_v}\right) +1 \right]
\end{equation}

which leads to the formation of two velocity shear layers centered at the two current sheets. The parameter $w_v$ sets the width of the velocity shear layer which was set to be equal to 1.0 for all our setups. As evident, the velocity shear  width $w_v$ and the magnetic shear width $w_B$ is identical for all the setups considered in our study. The quantity $v_{\rm s}$ determines the strength of the shear speed, which in our case, was varied from zero to the Alfv\'en speed of the system. An additional run with a super-Alfv\'enic shear ($v_{\rm s}/v_{\rm A}= 1.25$) was also performed for comparison. The names of the setups with various shear speeds are tabulated in table \ref{tab:setups}. An initial sinusoidal perturbation in the magnetic field containing a single wavelength was applied to the $B_y$ component to break the initial equilibrium condition. The mathematical form of the perturbation is described by eqn. (\ref{eqn:perturbation}). 

\begin{equation}\label{eqn:perturbation}
\begin{split}
B_y = 0.006 \times \rm{cos}\left( \frac{2\pi x}{128}\right) \rm{sin}^{2}\left( \frac{2\pi y}{64}\right)
\end{split}
\end{equation}

The corresponding $B_x$ component of the perturbation was derived from the constraint that $\nabla \cdot \mathbf{B}= 0$.  The flux computation has been done using the HLLD (Harten-Lax-van-Leer, `D' stands for discontinuities) Riemann solver along with a Vanleer Flux Limiter and the condition for $\nabla\cdot \mathbf{B}= 0$ is ensured by the Hyperbolic Divergence cleaning method\cite{Dedner_2002}. The resistivity is treated using a second order accurate explicit multistage Runge-Kutta Legendre time stepping scheme\cite{vaidya_RKL}.

\begin{table}
\caption{\label{tab:setups} Setup names for various shear speeds.}
\begin{ruledtabular}
\begin{tabular}{ccccccc}
  Shear ($v_{\rm s}/v_{\rm A}$) & 0.0 & 0.25 & 0.5 & 0.75 & 1.0 &1.25\\
\hline
 Setup Name (Fluid) & S0 & S25 & S50 & S75 & S100 & S125\\
 \hline
 Setup Name (Particles) & SP0 & - & - & SP75 & - & -
\end{tabular}
\end{ruledtabular}
\end{table}

\subsection{Domain and Resolution}

The spatial domain of the simulation is confined within $-L_{\rm x} \leq x \leq L_{\rm x}$ and $-L_{\rm y} \leq x \leq L_{\rm y}$. Periodic boundary conditions were applied along the x-boundaries and the y-boundaries are set to be reflective. In all our simulations, $L_{\rm x}= 64.0$ and  $L_{\rm y}$ is set at 96.0 to ensure that the boundaries are sufficiently far away from the current sheets in the y-direction and does not have any significant impact on the dynamics of the system. The simulation was carried out in a uniformly spaced Cartesian grid with a grid spacing of 0.1 in both, the x and y directions resulting in a $1280\times 1920$ grid cell layout. We performed resolution convergence tests with 1.5 times and twice the standard resolution employed here and the evolution pattern was very similar with respect to the initial phases and the onset of the plasmoid instability. Some deviation in the later stages is expected due to the inherent emergence of turbulence in plasmoid dominated systems. It is, however, important to note that the final relaxed states of the systems  were qualitatively similar even for the setups with higher resolution. The appendix contains a more detailed discussion on the convergence tests.

\subsection{Particles setup}

To study the effects of a shear flow from the perspective of particle acceleration during the explosive phase of the magnetic reconnection process, we have also studied the time evolution of test particles in our domain for various runs. The particle module in the PLUTO code describes the dynamical interaction between thermal plasma and a population of non-thermal, collisionless, charged particles called "CR Particles" \cite{Particle_PLUTO}. The background fluid component is evolved by shock capturing MHD methods as mentioned in Section \ref{Fluid Setup}, whereas, the charged particles are treated kinetically using the conventional PIC approach. The overall method is an MHD-PIC formalism introduced by Bai. et. al\cite{Bai_2015_MHDPIC}. The CR particles are defined in terms of their position coordinates ($\textbf{x}_{\rm p}$) and velocities ($\textbf{u}_{\rm p}$) which in turn are governed by the standard equations of motion given below:

\begin{equation}
\begin{split}
\frac{d \textbf{x}_{\rm p}}{d t} &= \frac{\textbf{u}_{\rm p}}{\gamma_{\rm p}}\\
\frac{d  \textbf{u}_{\rm p}}{d t} &= \alpha_{\rm p} \left(c\textbf{E} + \frac{\textbf{u}_{\rm p}}{\gamma_{\rm p}} \times \textbf{B} \right)
\end{split}
\end{equation}

where $\gamma_{\rm p}= \sqrt{1+(\mathbf{u}_{\rm p}/\rm{C})}$ is the Lorentz factor wherein `C' is the artificial speed of light that was set to be equal to 100$v_{\rm A}$ in the code. Such a parametrization allows us to study relativistic particle motion in an underlying non-relativistic fluid. $\alpha_{\rm p}= e/mc$ is the dimensionless CR charge to mass ratio which has a value of unity in our setup corresponding to protons. This choice of $\alpha_{\rm p}$ essentially sets the ion skin depth $c/\omega_{\rm pi}$ to be of unit length in the domain which renders the resolution employed by us for the particles to be in accordance as the MHD-PIC formalism dictates that resolving only the Larmor-scale is sufficient during the time integration of the particles. Indeed the ratio of typical ion gyroradii and the ion skin depth in the heliosphere varies from about $\sim 10^{-1}$ in the solar corona\cite{bingham_2004} to $\sim 10^0$ in the solar wind at 1AU\cite{verscharen_2019}. Recent laboratory plasma experiments have also shown that the neutral-sheet thickness in the field reversal region has a value $\sim 0.4 c/\omega_{\rm pi}$ which renders our choice of $\alpha_{\rm p}$ to be well suited with respect to the gradually thinning current sheet layers in our system having an initial width of $c/\omega_{\rm pi}$\cite{yamada_2000}. The electric and magnetic fields, \textbf{E} and \textbf{B} are interpolated from the magnetized grid to the particle positions using the Triangular-Shape-Cloud (TSC) interpolation approach (see [\onlinecite{Particle_PLUTO}] for more details). The feedback of the particle motions to the grid has not been included in this study. The particle positions and velocities are evolved using a standard relativistic Boris-pusher algorithm. A global time stepping scheme based on the fastest moving particle in the domain is employed for the time integration which ensures that no particle travels more than $N_{\rm max}=1.8$ zones (3-D grid cell) in a single step and that the Larmor scale is resolved in more than one cycle \cite{Particle_PLUTO}. We inject a total of $4\times 10^6$ particles in a region bounded by $-64\leq x \leq 64 $, $-40\leq y \leq 40 $ and $-10\leq z \leq 10 $. Since the particles have 3 degrees of freedom, we have extended the same 2D setup in the third dimension with a small thickness along the z-coordinate ($-L_{\rm z} \leq z \leq L_{\rm z}$) with $L_{\rm z}$= 10.0 with periodic boundaries. This essentially stretches the exact 2D slice of the fluid simulations in the direction of the third coordinate which in turn deems the fluid quantities to be invariant along the z-coordinate. While the length scales are normalised to c/$\omega_{\rm pi}$, the time is given in the units of the inverse cyclotron frequency $\Omega^{-1}= c/(\omega_{\rm pi} v_{\rm A})$. Therefore, the evolution times for the MHD-PIC runs have been expressed in terms of $\Omega t^{\prime}$ which is equivalent to the fluid evolution time `$t$' normalised to $\tau_A$. The particle positions are initialized randomly within the injection region with a Gaussian distribution of initial velocities having a mean of zero and a standard deviation of 0.1$v_{\rm A}$. This statistically means that $\sim 68 \%$ of the particles have an initial speed in the range of 0$v_{\rm A}$ to 0.173$v_{\rm A}$ with much more of the particles having a speed close to the lower limit. This choice was encouraged by the fact the typical ion speeds in the solar corona\cite{akramov_baty} is of the order of $10^{-3} v_{\rm A}$ to $10^{-4} v_{\rm A}$. The boundary conditions for the particles are exactly the same as the fluid boundaries, i.e periodic along the x and z-boundaries and reflective along the y-boundaries.

\section{\label{Results}Results:\protect}

\subsection{General Evolution Characteristics\label{Results:General_evolution}}

\begin{figure*}[ht]
    \centering
    \includegraphics[width=1.0\textwidth]{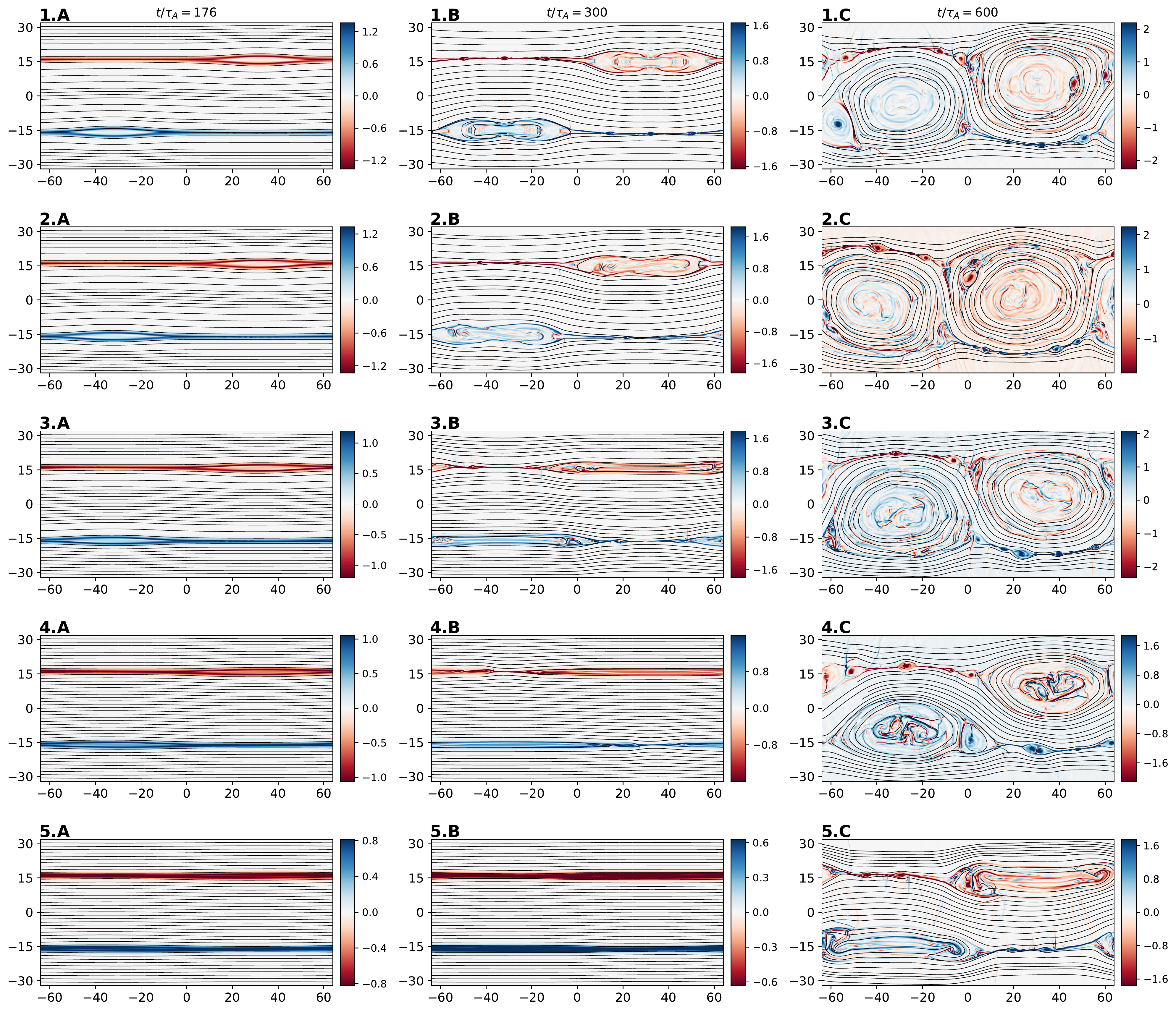}
    \caption{Time evolution of the systems for various shear speeds and three different times. The first , second and the third columns are snapshots at $t/\tau_{\rm A} = 175$, $t/\tau_{\rm A} = 300$ and $t/\tau_{\rm A} = 600$ respectively while the rows represent the setups S0 (0.0$v_{\rm A}$), S25 (0.25$v_{\rm A}$), S50 (0.5$v_{\rm A}$), S75 (0.75$v_{\rm A}$) and S100 (1.0$v_{\rm A}$) respectively. The background colors represent the out of and into the plane current density \textbf{J} which is over-plotted with magnetic field streamlines. The subplots show the entire x-span of the domain ($-64\leq x \leq 64 $) whereas they have been truncated in the y-direction to $-32\leq y \leq 32 $.} A multimedia file showcasing the complete evolution of the S0 setup is available in the online version (Multimedia View).
    \label{fig:evolution_all}
\end{figure*}

 We first report the results of the MHD-fluid setup with no shear flow (S0) across the current sheets as a benchmark to compare the differences in the evolution for the cases with a non-zero shear. During the initial phases of the evolution, the long wavelength perturbation applied to the current sheets disrupt the initial equilibrium configuration of the system. The mode of instability corresponding to the single wavenumber along the x-direction applied in the perturbation (eqn. \ref{eqn:perturbation}) grows in the form of two magnetic islands centered at $x \sim 32.0$ for the upper current sheet and $x \sim -32.0$ for the lower current sheet as can be seen in panel 1.A  of Figure \ref{fig:evolution_all}. The primary magnetic islands assume an anti-symmetric configuration due to the dominance of the anti-symmetric mode over the symmetric mode \cite{sym_antisym/1.860337}.
 
 Earlier observations coupled with simulations have shown that the evolution of a double current sheet configuration can be divided into four separate phases, an early growth phase, a transition phase, a fast growth phase and an overall relaxation or decay phase which can be identified based on the evolution of the kinetic energy of the system with time \cite{wangetal2007, chang1996}.
 
 \begin{figure}[ht]
    \centering
    \includegraphics[width=0.48\textwidth]{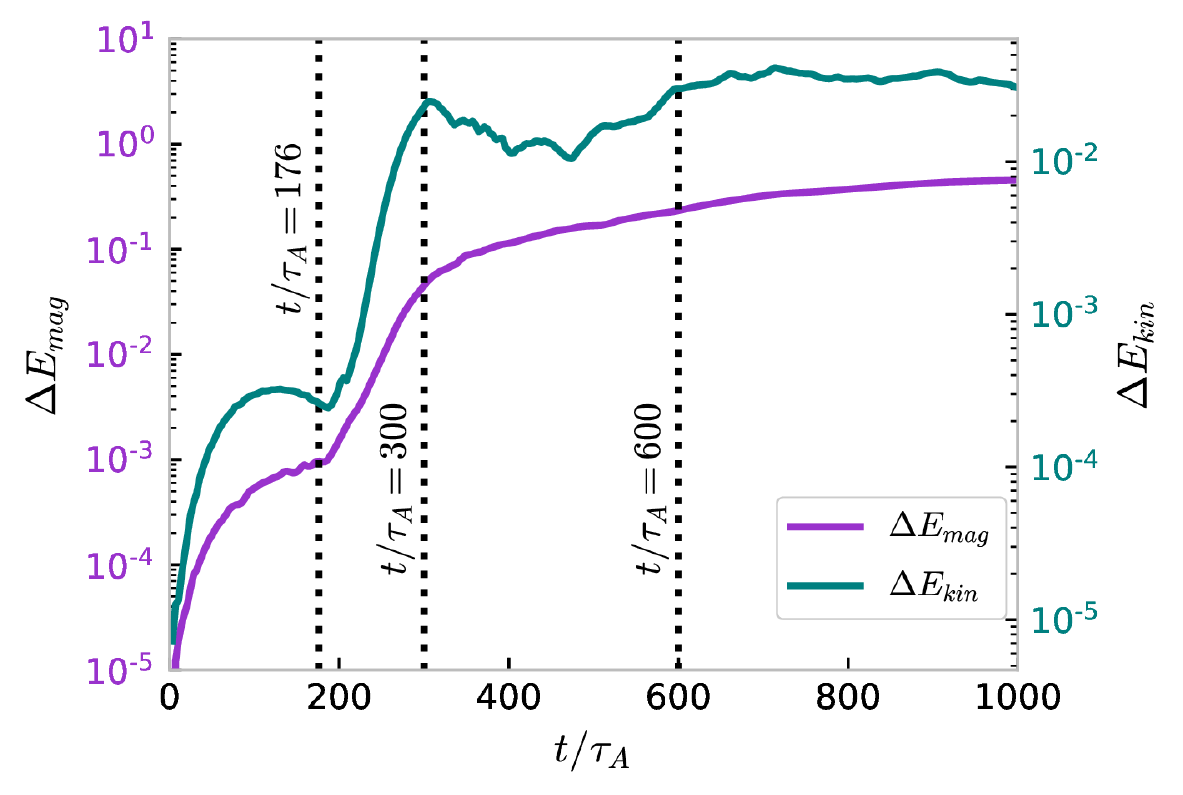}
    \caption{Plot showing the magnetic energy change and the kinetic energy change with time for the S0 system. The change in the magnetic energy is determined as $\Delta E_{\rm mag}= E_{\rm mag(t=0)} - E_{\rm mag(t)}$ and the change in the kinetic energy is determined as $\Delta E_{\rm kin}= E_{\rm kin(t)} - E_{\rm kin(t=0)}$. The vertical dotted lines correspond to the snapshots shown in the first row of Figure \ref{fig:evolution_all}.} 
    \label{fig:magen_kinen_sh0}
\end{figure}

\begin{figure}[ht]
    \centering
    \includegraphics[width=0.48\textwidth]{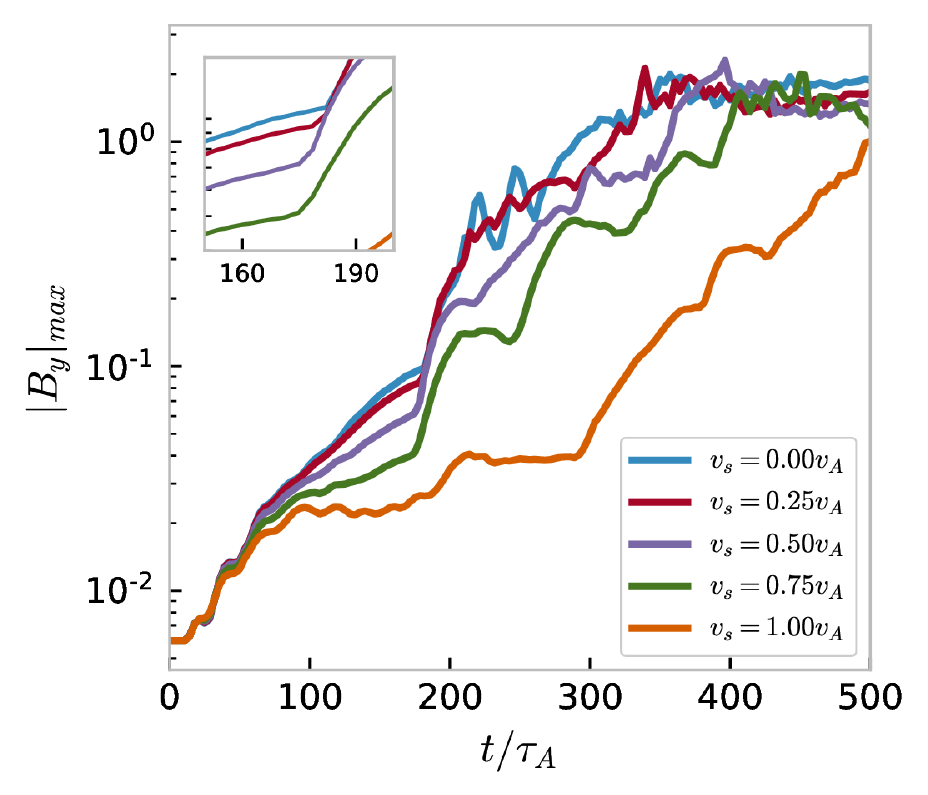}
    \caption{A plot showing the comparison of the $\mid B_y\mid_{max}$ component in the domain as a function of time for various shear speeds across the current sheet layers.}
    \label{fig:bymax_for_all_shears}
\end{figure}

\begin{figure}[ht]
    \centering
    \includegraphics[width=0.48\textwidth]{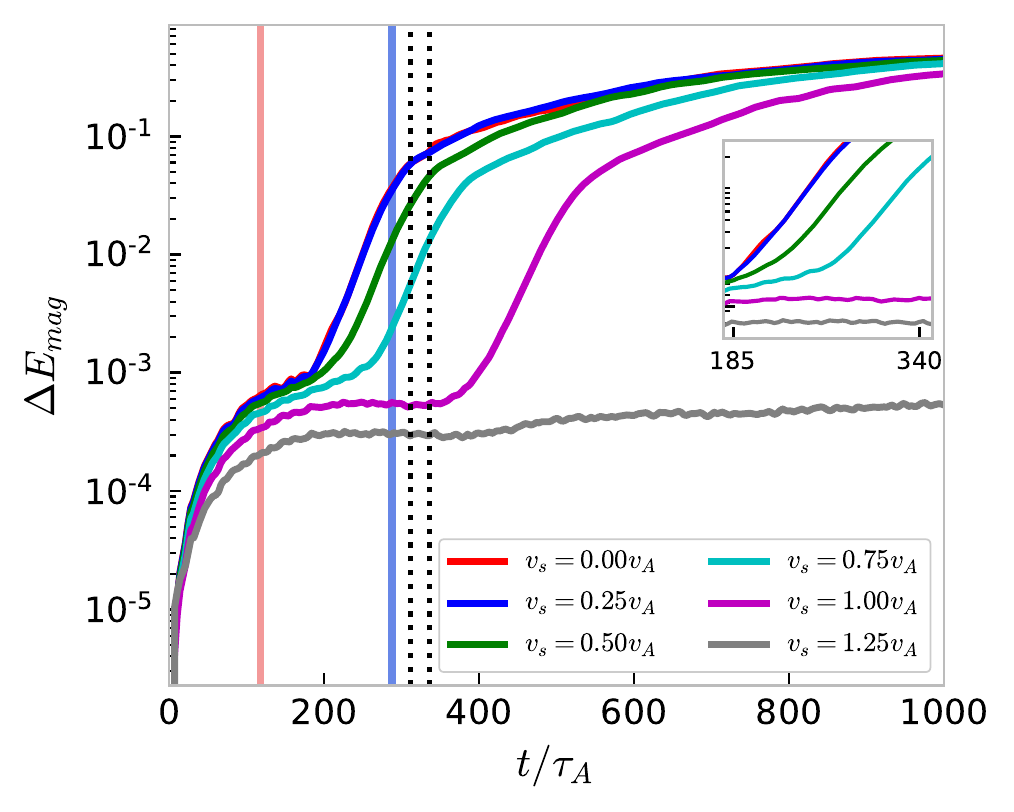}
    \caption{A comparison of the change in magnetic energy density $\Delta E_{\rm mag} = E_{\rm mag(t=0)} - E_{\rm mag(t)}$ in the domain with respect to time for various shear speeds. The inset shows a zoomed in region during the beginning of the non-linear plasmoid instability phase for all shear speeds. We have also included the results from an auxiliary run with super-Alfv\'enic ($v_{\rm s}/v_{\rm A} =1.25$) shear speed for comparison. {The red and blue vertical strips correspond to the time durations $t/\tau_{\rm A} = 113$ to 123 and $t/\tau_{\rm A} = 283$ to 293 respectively whereas the two dotted vertical lines correspond to $t/\tau_{\rm A} = 311$ and $t/\tau_{\rm A} = 336$ respectively (see section \ref{Sec:Recon_Rate} for details)}.}
    \label{fig:mag_en_for_diff_shears}. 
\end{figure}

 Figure \ref{fig:magen_kinen_sh0} shows the time evolution of the magnetic energy ($\Delta E_{\rm mag}$) and the kinetic energy ($\Delta E_{\rm kin}$) of the S0 system. The kinetic energy starts to increase near linearly, starting at  $t/\tau_{\rm A} \sim 12$ up to $t/\tau_{\rm A} \sim 70$. This is also around the same time when the two large magnetic islands have just started to form and are visually distinguishable in the domain. The change in magnetic energy also shows a similar linearly decreasing trend (Note that the figure \ref{fig:magen_kinen_sh0} plots the difference in the magnetic energy w.r.t its initial value in the domain). The reconnection proceeds as can be seen from the profile of $\Delta E_{\rm mag}$ which keeps decreasing. The increase in the domain averaged kinetic energy tends to stabilize in between $t/ \tau_{\rm A} \sim 100$ to $t/\tau_{\rm A} \sim 150$ followed by a mild decrease between $t/\tau_{\rm A} \sim 150$ to $t/\tau_{\rm A} \sim 175$ as the counter-streaming reconnection exhausts mix inside the primary magnetic islands. We identify this rather brief phase as the Rutherford regime where the reconnection proceeds in a purely diffusive timescale. We note that the duration of this phase is smaller than what has been reported in previous studies \cite{akramov_baty}, the reason for this is the fact that the duration of the Rutherford regime of magnetic reconnection tends to decrease with an increase in the length of the domain \cite{Janvier2011}. To check the consistency of the growth rate during the early phases corresponding to the applied mode of perturbation, we have fitted an exponential curve to the early phases of the time evolution of the maximum value of the $B_y$ component ($|B_y|_{max}$) during $t/\tau_{\rm A} \sim 65$ to $t/\tau_{\rm A} \sim 165$. The value of $|B_y|_{max}$ can be considered as an arbitrary measure of the instability magnitude for such systems\cite{Baty_2017, akramov_baty}. Indeed the range can be fitted to an exponential of the form of $|B_y|_{max} \propto \rm exp(0.013 t/\tau_A)$. We thus deduce the growth rate to be 0.013 when normalized to the Alfv\'en transit time scale $\tau_A$. This is well in accordance with the theoretical value of 0.014 calculated from the linear theory of DTMs having dual hyperbolic tangent profiles as described in Pucci et. al. (2018) \cite{Pucci_2018}. During this entire period the primary magnetic islands have grown in size due to the inflow of the reconnection exhausts from the initial X-points. The increase in size of the magnetic islands eventually leads to a coupling between the two current sheet layers and the dynamics of the evolution transitions from a single current sheet evolution to that similar to a double tearing mode (DTM) when the coupling becomes significant. An important parameter to characterize the current layers is the inverse aspect ratio given by $\delta /L$ where $\delta$ is the thickness of the current sheets and $L$ is the length. The secondary current sheets (portion of the current sheets outside the large magnetic islands) during their evolution initially tend to approach towards the steady state Sweet-Parker inverse aspect ratio $\delta_{\rm SP}/L$ given as $\delta_{\rm SP} /L \sim S^{-0.5}$. This results in a thinning of the current sheets which increases their aspect ratio\cite{sweet_parker1, sweet_parker_2, loureiro_uzdensky_rec, Comisso_2017}. The increase in the aspect ratio continues until it reaches a critical threshold value. Thereafter at $t/\tau_{\rm A} \sim 175$ a portion of the current sheets becomes unstable due to the onset of the plasmoid instability\cite{hossienpour_plasmoid_shear}. We have measured the length and the thickness of the secondary current sheets by determining the FWHM of the current density profile along $y= \pm 16$ (for the length) and along $x=\pm 32$ (for thickness) just before the disruption of the sheets. We have thus computed an average (over the two layers) value of $L= 62.1$ and $\delta= 0.37$ for the secondary current sheets just before the onset of the plasmoid instability. Recent theoretical studies on large aspect ratio current sheets have revealed that the Sweet-Parker inverse aspect ratio  $\delta /L \sim S^{-0.5}$ leads to  diverging growth rates at high Lundquist number values\cite{Huang_2017}. Pucci and Velli (2014)\cite{Pucci_2013} thus suggested that the current sheets are disrupted before they reach the Sweet-Parker aspect ratio. For a general situation where the current sheet inverse aspect ratio has the dependence $\delta /L \sim S^{-\alpha}$, they concluded that the current sheet disruption occurs at $\alpha = 1/3$. The value of the Lundquist number for the secondary current sheets formed in our simulation (S0 setup) is $\sim 4.4 \times 10^ 6$ which is well beyond the critical threshold of $\sim 3\times 10^4$  reported by Bhattacharjee et. al.(2009)\cite{Bhattacharjee_2009}. The inverse aspect ratio $\delta/L$ just before disruption of the current sheets was calculated to be $5.95 \times 10^{-3}$ which is remarkably close to the critical value of $6.1 \times 10^{-3}$ derived from the aspect ratio scaling given by Pucci and Velli (2014). Beyond this aspect ratio, the current sheet violently disrupts and fragments into multiple small plasmoids. Thus, at $t/\tau_{\rm A} \sim 175$ a portion of the current sheets becomes unstable due to the onset of the plasmoid instability\cite{hossienpour_plasmoid_shear}. The formation of such plasmoids in multiple current sheet systems has previously been reported in literature for large aspect ratio current sheets\cite{nemati_2016}. Comisso et.al. (2017) have presented a theoretical framework and correspondingly worked out an expression for the growth rate and the dominant wavenumber of plasmoid instability in thin current sheets\cite{Comisso_2017}. We have once again calculated the growth rate during the early phases of the plasmoid instability ($t/\tau_{\rm A} \sim 180$ to $t/\tau_{\rm A} \sim 220$) in a similar manner as mentioned before and arrived at a value of $\gamma \tau_{\rm A} \sim 4.3 \times 10^{-2}$. This is of the same order as the theoretical approximate of $\gamma \tau_{\rm A} \sim 1.7 \times 10^{-2}$ obtained by setting the $\mathcal{O}(1)$ constant $c_{\gamma} =1$ in eqn. (17) of Comisso et. al. (2017)\cite{Comisso_2017}. The theoretical prediction of the dominant wavenumber infers the number of plasmoids during the very early phases of the plasmoid instability to be $\sim 13$. This is also in very close agreement to the 16 initial plasmoids observed during the early phases of plasmoid instability in each of our secondary current sheets. The onset of the plasmoid instability occurs at nearly the same time for all the setups as can be seen from the spikes of $\mid B_{y}\mid _{\rm max}$ in the inset region of Figure \ref{fig:bymax_for_all_shears}. We also observe a very marginal advancement of the onset time for increasing shear. Studies on the linear and non-linear phases of the plasmoid instability of thin current sheets have demonstrated that the non-linear phase of the plasmoid instability (steep slopes in the $\Delta E_{\rm mag}$ curves in Figure \ref{fig:mag_en_for_diff_shears}) occurs when the width of the plasmoids exceed the current sheet width\cite{Comisso_2017}. The presence of a shear for the setups S25, S50 and S75 suppresses the growth of the plasmoid instability during its initial phases of evolution by choking the tendency of the plasmoids to grow larger than the current sheet width, and thus there is a slight delay in the onset of the explosive non-linear phase for the setups with increasing shear. It has also been previously highlighted by Janvier et. al. (2011) that the time evolution of a DTM consists of an island structure driven non-linear instability which develops to exhaust the free energy associated with the deformation of the magnetic islands. Simply stated, the large magnetic islands tend to drive magnetic flux towards the X- points located above them resulting in a fast-reconnection phase. The instability criteria states that the non linear phase is triggered if the ratio of the domain size in the periodic direction (x-coordinate) and the half separation between the current sheets, i.e $2L_{\rm x}/l \geq 6$ \cite{akramov_baty, Janvier2011}. For our setups, the value of  $ 2L_{\rm x}/l\sim 8$ and thus we expect that our setups will exhibit this structure driven non-linear instability. To identify the effects of this structure driven instability in the evolution of our double current sheet system, we have compared the evolution of the S0 setup with two independent single tearing modes(STMs) (results not shown) to precisely identify the time when the current sheets evolve from two independent STMs to a state where the coupling between the layers is significant as expected from a DTM. {The STMs have exactly the same initial properties (aspect ratio, perturbation, resistivity, etc.) and thus the difference in evolution, if any, would arise from the interaction of the two current sheets of the DTM. The plasmoid instability is also excited in the STMs as the onset of the plasmoid instability is solely by virtue of the thinning of the current sheets. However, we find that the DTM configuration deviates from the independent STMs evolution pattern just after the beginning of the non-linear phase of the plasmoid instability when the size of the primary magnetic islands increase rapidly}. {We thus assert that the explosive non-linear phase of the evolution of the $\Delta E_{\rm mag}$ is due to a combination of the plasmoid instability, and a flux drive characteristic to the structure driven instability in DTMs.} This phase can be distinguished from the steep slope of the magnetic energy evolution between $t/\tau_{\rm A} \sim 175$ and $t/\tau_{\rm A} \sim 300$. During this period, the thinned current sheets break down into several small plasmoids that grow in size and gradually feed into the magnetic islands. The domain averaged kinetic energy density of the system during this explosive phase also shows a steep increase that pushes its value higher by almost two orders of magnitude. We observe that multiple X-points have formed in the plasmoid unstable region. As time progresses, plasmoids are continuously produced and they advect along the reconnection jets towards either side of the two large magnetic islands along the current sheets and merge with them which leads to a gradual increase in the size of the islands. Also, at $t/\tau_{\rm A} \sim  300$ the primary magnetic islands have grown large enough that the upper current sheet is visibly bent due to an increase in size of the lower magnetic island and vice versa. This bend can be seen in the panel 1.B of Figure \ref{fig:evolution_all} which is plotted around the same time. Due to this curving of the current sheets, there appears a slight displacement in the y-direction (exhausts entering the plasmoids are not head-on) between the counterstreaming small-scale outflow exhausts entering the miniature plasmoids that are forming in the plasmoid unstable region which leads to the formation of fluid vortices inside these ejected plasmoids. The continuous feeding of the plasmoids gradually increases the size of the primary magnetic islands. We name the two largest magnetic islands (one for each current sheet) in our domain as the `monster-plasmoids' and hereafter the terms `largest magnetic islands' and `monster plasmoids' are used synchronously with each other. At the later stages, these monster-plasmoids become large enough and structurally deform to a triangular shape similar to what has been reported in previous studies\cite{akramov_baty, Janvier2011}. Further in the evolution, when the size of the two monster-plasmoids become comparable to the separation of the current sheets, the field lines that are a part of the outer edge of the largest magnetic islands start to reconnect at the X-points of the other current sheet (Panel 1.C of Figure. \ref{fig:evolution_all}).{The duration when the primary magnetic islands are large in size (comparable to the current sheet separation) is generally expected to be the phase where the secondary structure driven instability is significantly dominating, however, the current sheets in our system is poplulated with multiple plasmoids during this period and this has the potential to render the DTM flux drive to be less efficient. This phenomenon, however, is responsible for the gradual increase in the $\Delta E_{kin}$ of the system during $t/\tau_{\rm A} \sim  500$ to 650 as seen from figure \ref{fig:magen_kinen_sh0}.} This interaction eventually leads to an exchange in the vertical position of the islands which finally coalesce together at $t \sim 660 $ and gradually disappear leaving behind a filamentary turbulent magnetic structure. We also note the occurrence of secondary reconnection current sheets during the merger of the plasmoids with the large magnetic islands(see bottom left of panel 1.C in Figure \ref{fig:evolution_all}), however, the formation and evolution of these secondary reconnection process does not have a significant effect on the overall dynamics. A multimedia file showing the complete evolution of the S0 system is provided with the online version of the manuscript (see Figure: \ref{fig:evolution_all}). Wu et. al. (2013) have performed studies on DTM in the presence of a shear flow to identify its effects on the onset of the fast-reconnection phase. They reported that there is a delay in the onset of the non-linear phase of the DTM with an increase in the magnitude of the shear flow velocity\cite{DTM_shear_wu}. We observe from our simulations that for the cases with increasing shear flow described below, the general evolution pattern is similarly delayed in time due to the analogous delay in onset of the fast reconnection phase. However their study differs from the one presented here from the standpoint that any quantification of the scaling of the reconnection rate with shear was not reported and the onset and evolution of the plasmoid instability was also absent in their simulations.
 
 For the S25 setup (with shear speed of 0.25$v_A$), the early time evolution of the system is very similar to the setup without a shear flow (S0). The formation and evolution of the initial magnetic islands and the onset of the plasmoid instability occurs at nearly the same time in the setups S0 and S25. It has been observed by Chen et. al. 1997 that the presence of a parallel shear flow reduces the thickness of the magnetic islands and stretches the islands to an elongated form. Our results also demonstrate a consistent behaviour whereby the size of the largest islands in the S25 setup in our simulations appear to be very slightly elongated when compared to the S0 setup. The same can also be seen upon careful observation of panel 2.A of Figure \ref{fig:evolution_all}.
 This trend continues for increasing shear speeds as will be seen later in this section. We also observe a displacement of the outflow exhausts with respect to each other along the y-direction inside the largest magnetic islands. This kind of displacement has the potential to form large scale Kelvin-Helmholtz vortices inside the monster plasmoids as has also been seen in previous studies employing high resolution Particle-In-Cell(PIC) simulations\cite{Huang_KH_inside_islands}. After the onset of the plasmoid instability, when the width of the plasmoids are comparable to the current sheet width, vortices start to form inside the plasmoids as they interact with the counter-streaming fluid flow along its upper and lower surfaces. We note that these vortices form much earlier than the vortices for the S0 setup, which is solely due to the bending of the current sheets and only starts when the current sheets are significantly curved. We note here that the velocity profile of the shear flow used in our simulations is similar to the magnetic field profile. The velocity shear width is also equal to the magnetic field shear width. Recently, a study on  the effect of varying the shear width of the velocity profile has been put forward by Hosseinpour et. al. in the context of the plasmoid instability\cite{hossienpour_plasmoid_shear}. It has been found that for extremely small shear flow width (< $w_{B}/3$), the system is stabilized against the plasmoid instability even for a sub-Alfv\'enic shear (0.65$v_{A}$). The plasmoid instability is again seen to develop when the shear width is increased (> $w_{B}/3$). The shear width profile in our setups are thus, in the later limit and the existence of plasmoid instability in our system is congruous with this study.
 
 For the case of the system S50, the initial magnetic islands are significantly thinner and elongated. The non linear phase of evolution with plasmoid instability feeding the initial islands to form large `monster-plasmoids'  visually looks similar, but a closer look at the evolution of the maximum value of the $B_y$ component (Figure: \ref{fig:bymax_for_all_shears}), show that the growth of the perturbations is suppressed significantly in the presence of a shear flow of this magnitude during the initial phases and also during the explosive phase. Hence, even though the growth and evolution pattern of the system proceeds along the same path, the explosive phase is slightly delayed in time by $\Delta t/\tau_{\rm A} \sim 50$. It is due to this suppression in the instability growth that the rate of change of the magnetic energy with time shows a flatter slope during the start of the non-linear phase for the case S50 as can be seen from Figure \ref{fig:mag_en_for_diff_shears}.
 
For the systems with even higher shear speeds, the initial phases, continue to follow the similar trend of suppressed instabilities resulting in a slower growth rate of magnetic energy dissipation. We also note that the half-width of the plasmoids ejected during the non linear phase decreases with an increase in shear speed.

The S100 setup passes through a slightly altered evolution process. The growth of the plasmoid instability is significantly delayed because the perturbations are severely choked in the presence of an Alfv\'enic shear and reconnection is highly suppressed during the initial phases. The system shows a tendency to form large scale primary magnetic islands but this is quickly destroyed by the shear flow as their width starts exceeding the current sheet width. It is at a much later time ($t/\tau_{\rm A} \sim 350$) that the current sheets become plasmoid unstable. A study on the non-linear plasmoid instability in the presence of a shear flow for a single current sheet system has been recently put forward by Hosseinpour et.al (2018) \cite{hossienpour_plasmoid_shear}. They report that their systems exhibit the highly non linear plasmoid instability phase even in the presence of an Alfv\'enic velocity shear\cite{hossienpour_plasmoid_shear}. Our simulation also shows a similar result for the double current sheet configuration. We observe the highly non linear plasmoid instability even in the presence of an Alfv\'enic shear flow, even though reconnection in the initial phases ($t \leq 350$) is highly suppressed. Once the plasmoid instability starts, the further evolution is similar to the cases with lesser shear speeds, i.e, the plasmoids feed into the large magnetic islands and finally result in a global merger. It is of interest to note that the onset of the plasmoid dominated fast-reconnection phase occurs at later times for higher shear speeds. The principal reasons behind this are as follows. Firstly, the growth of the perturbations is increasingly suppressed for higher shear speeds leading to a slower transition into the highly non-linear phases. Secondly, the transition to a  fast nonlinear regime is due to a combination of plasmoid instability and structure driven instability. The later only occurs when the primary magnetic islands assume a certain critical deformation \cite{Baty_2017, Janvier2011}. This critical deformation depends on (a)the reconnection rate which feeds the reconnection exhausts into the magnetic islands, (b) the half widths of the primary magnetic islands which tend to be smaller with increasing shear. As an auxiliary reference, we have also shown the energetics of a super-Alfv\'enic shear in Figure \ref{fig:mag_en_for_diff_shears} marked by the grey line. This confirms that there is no explosive reconnection {as well as plasmoid instability} for super-Alfv\'enic shears. We  note here that irrespective of the shear speed, we observe the presence of small scale Petschek like shocks emerging from the plasmoid dominated region. The presence of such shock structures has been found to facilitate fast  reconnection in plasmoid dominated systems\cite{shibayama_2016}. A more detailed analysis is required to quantify the effects of these small scale shocklets in governing fast reconnection in our simulations which is beyond the scope of the current study. We also note here that the maximum shear speed considered in our study is 1.25$v_{\rm A}$. This corresponds to an Alfv\'enic Mach number ($M_A$) that is within the stability criteria $M_A \leq 2$ for the KH instability to grow\cite{Miura_1982}. Furthermore, compressibility of the plasma is also known to have a stabilizing effect on the KH modes\cite{hossienpour_plasmoid_shear}. Thus, we find that our system is stabilized against the KH instability across the magnetic field reversal layers for all the velocity shears considered in this study. KH instabilities, however, are important from the context of magnetic reconnection as the formation of the KH vortices have the potential to squeeze the field lines together thereby establishing vortex induced reconnection (VIR). VIR can occur even in the absence of magnetic field line reversals. The study of the synergy between the KH instability and reconnection has been of interest in literature\cite{Pu_1990, Ma_2014, Labelehemer_1988, Keppens_1999, Erikson_2012} but is beyond the scope of the present study. Next we move on to the analysis of the reconnection rate at various times during the evolution.

\subsection{\label{Sec:Recon_Rate}Reconnection Rate:\protect}
Theories and simulations have confirmed that the reconnection rate is suppressed in the presence of a shear flow and is completely switched-off if the shear flow is super-Alfv\'enic\cite{Labellehamer_1994, mitchell_kan_1978}. Cassak and Otto (2011) drew a strong analogy between this switch-off effect and the results showing that reconnection at the magnetopause is suppressed when the diamagnetic drift is super-Alfv\'enic\cite{Swisdak_2003} which has been confirmed by observations in the solar wind\cite{Phan_2010}. They also gave a quantitative expression for the suppression of the reconnection rate with an increasing shear flow\cite{Cassak_otto_2011}. It is thus known that for a single current sheet, the reconnection rate decreases with an increasing velocity shear ($v_{\rm s}$) following the relation:

\begin{equation}\label{eqn:recon_shear}
    E= E_0 \left(1- \frac{v_{\rm s}^2}{v_{\rm A}^2} \right)
\end{equation}

where $E_0$ is the reconnection rate without any shear. To test this scaling for a double layered current sheet during the early phase and the explosive secondary instability phase, we have measured the reconnection rate of the coupled current sheet system. The reconnection rate is given as the sum of the average values of the magnitude of the electric field $E_{\rm z}$ in two thin strips ($\Delta y= 2.0$) along the two current sheets. This is normalized to $B_0 v_{\rm A}$ and its mathematical formulation is given below:

\begin{equation}
    E_{\rm rec}= \frac{1}{B_{0} v_{\rm A}} \left(\left< \left| \frac{\iint_{y_{1}}^{y_{2}} E_{\rm z}dxdy}{\iint_{y_{1}}^{y_{2}} dxdy} \right| \right> + \left< \left|\frac{\iint_{-y_{1}}^{-y_{2}} E_{\rm z}dxdy}{\iint_{-y_{1}}^{-y_{2}} dxdy} \right|\right>\right).
\end{equation}

\begin{figure}[ht]
    \centering
    \includegraphics[width=0.48\textwidth]{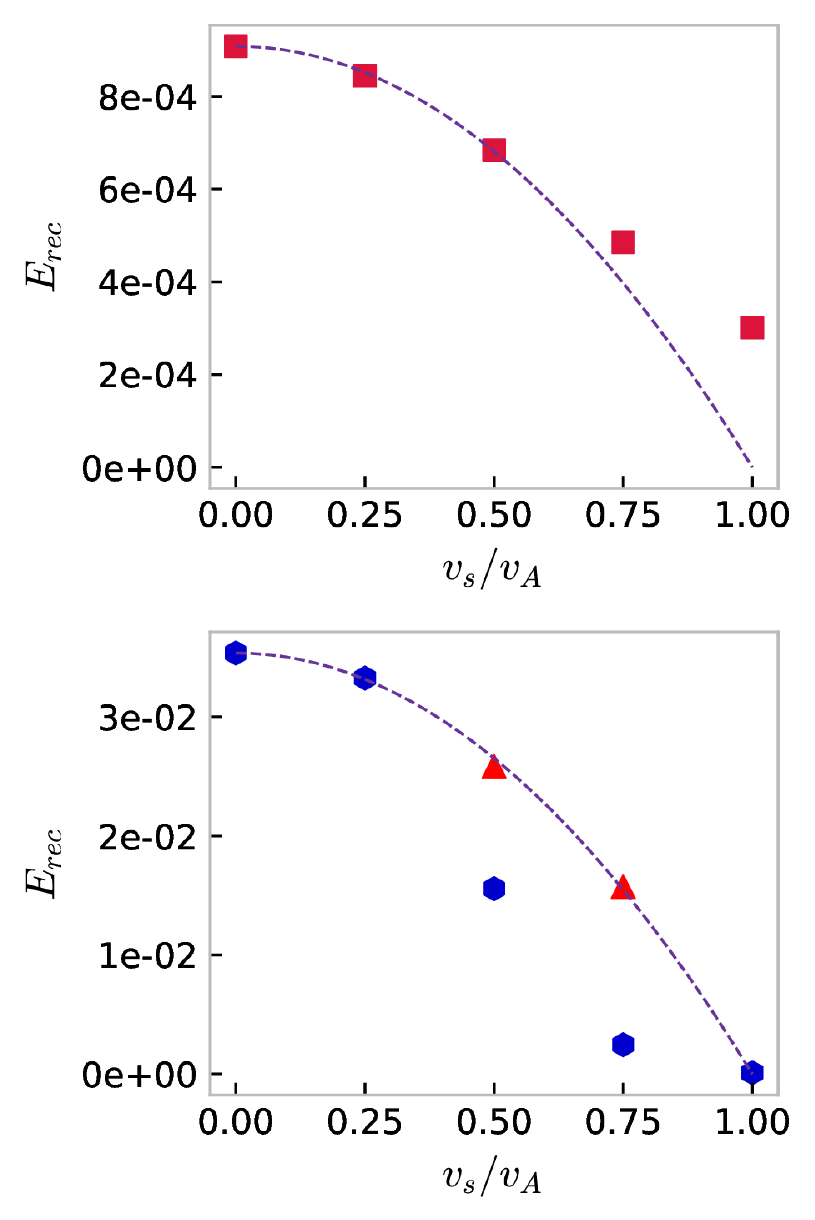}
    \caption{A plot of the variation of the reconnection rate with shear for two different time spans. Top panel shows the reconnection rate averaged between $t/\tau_{\rm A}$=113 to $t/\tau_{\rm A}$=123 (early phase) whereas the bottom panel shows the reconnection rate averaged between $t/\tau_{\rm A}$= 283 to $t/\tau_{\rm A}$=293 (explosive phase). The dashed lines represent the theoretical curves and the red squares and the blue hexagons are representative of the values obtained from the simulations (details in the text).}
    \label{fig:rec_rate_vs_shear}
\end{figure}

Here $y_1= 15$ and $y_2=17$ enclosing the current sheets. Figure \ref{fig:rec_rate_vs_shear} shows two panels plotting the variation of the reconnection rate averaged over two different time spans for our setups. The formulation that describes the electric field (equation \ref{elec_F}) consists of a convective and a resistive term and the reconnection rate in based on the fact that the contribution to the electric field is given by the resistive term along the middle of the current sheet width, and by the convective term elsewhere. Since we are averaging over a thin strip enclosing the current sheets, both of these terms tend to be equally important in measuring the reconnection rate. Also, due to the highly time varying nature of the systems exhibiting plasmoid dominated reconnection, it is difficult to achieve a stationary state over which the reconnection rate stabilizes. Hosseinpour et. al (2018)\cite{hossienpour_plasmoid_shear} have described the reconnection rate for a time varying plasmoid unstable current sheet by averaging the reconnection rate over a certain time span (typically 10 $\tau_A$). To mitigate this issue of the presence of a non-stationary reconnection rate, we have therefore employed a similar procedure and have averaged the reconnection rate over a time span of 10 $\tau_A$ for both the early phases and the fast non-linear phases to describe the scalings.

The top panel of Figure \ref{fig:rec_rate_vs_shear} shows the reconnection rate averaged over $t/\tau_{\rm A}= 113$ to $t/\tau_{\rm A}= 123$, i.e. during the early phases of the evolution {(all the characteristic times related to the reconnection rate has also been plotted in figure \ref{fig:mag_en_for_diff_shears})}. The theoretical curve {plotted using equation (\ref{eqn:recon_shear})} matches the results obtained from the simulation quite well during this period. During this period the system evolves as two separate current sheets and there is no significant interaction between the layers. We note that the reconnection rate shows a slight deviation from the theoretical curves for the S100 system. {Even though it is difficult to isolate the exact cause of this slight deviation, it may be due to the presence of a  {variable} Alfv\'en speed inside current sheet width attributed to the variation in $B_x$ and $\rho$ that is required for achieving the initial equilibrium. Another possible reason for the deviation could be the measurement error arising from the interaction of the higher shear flows and the perturbation provided to start the reconnection, in calculating the reconnection rate, which is inherently very small during the early phase. Such measurement errors become insignificant when the reconnection rate eventually increases. A small deviation from the theoretical rate at high shear speeds has also been reported in literature by previous studies concerning the scaling of the reconnection rate with shear for single current sheets\cite{Cassak_otto_2011, cassak_2011}.}
The blue hexagons in the bottom panel of Figure \ref{fig:rec_rate_vs_shear} shows the reconnection rate averaged during a much later time (explosive phase) of evolution in the interval $t/\tau_{\rm A}= 283$ to $t/\tau_{\rm A}=293$. There is a significant deviation from the theoretical curve for all shear speeds above $v_{\rm s}/v_{\rm A} =0.25$. This is due to the fact that in such a configuration similar to a DTM, the reconnecting layers have a feedback effect on each other. Since the islands form in an anti-symmetric configuration, an increase in the size of the magnetic islands tend to enhance the magnetic flux that is pushed towards the reconnecting X-points located above them. For the case of low shear speeds, the width of the primary magnetic islands tend to be larger in the Y-direction than those with higher shear speeds as can be seen from the middle row in figure \ref{fig:evolution_all}. This increase in size results in an efficient coupling between the two layers which in turn translates to more magnetic flux being driven towards the X-points in the region between the two islands which significantly boosts the reconnection rate for low shear speeds. For higher shear speeds, the flattened structure of the primary magnetic islands deems that the coupling is weaker which results in a remarkably smaller reconnection rate for such cases. As a test for this, to isolate the effects of the higher shear speeds on the dynamics of the coupled tearing mode during the explosive phase, we have calculated the reconnection rate for the systems S50 and S75 at later times ($t/\tau_{\rm A}= 311$ and $t/\tau_{\rm A}= 336$ respectively) when the width of the primary magnetic islands for these two systems are similar to the width of the island ($\Delta y \sim 7.8$) from the SP0 setup when the initial measurement of the reconnection rate was done (between $t/\tau_{\rm A}= 283$ to $t/\tau_{\rm A}= 293$). The results are plotted in red triangles in the right panel of Figure \ref{fig:rec_rate_vs_shear}.  {This shows that when the feedback effect arising due to the structure of the magnetic islands, wherein the bulge of the large magnetic islands tend to push magnetic flux towards the X-points above/below them, are taken into account (by calculating the reconnection rate when the island sizes are similar), the scaling of the reconnection rate with shear given by equation \ref{eqn:recon_shear} holds true even during the non-linear phases of the evolution of the DTM.}

\subsection{\label{Sec:Particle}Particle Energetics:\protect}

As per the particle setup mentioned in section \ref{Setup}, we have integrated the test particles  {(protons)} trajectories to see the impact of the explosive reconnection phase on the particle energetics and their variation with the presence of a shear speed. We probe into the effects that the presence of a velocity shear has on the particle acceleration process by injecting the particles in two different fluid setups, S0 and S75. We correspondingly name the fluid + particle setups SP0 and SP75. A total of 4 million test particles with a Gaussian velocity distribution having zero mean and a standard deviation of 0.1$v_{\rm A}$ along each coordinate are employed. To isolate the effects of the explosive phase in energizing the particles, we injected the particles into the domain just before the start of the highly non-linear reconnection phase for the setups S0 and S75 as identified from Figure \ref{fig:mag_en_for_diff_shears}. For the SP0 setup, this corresponds to an injection time of $\Omega t^{\prime}= 172.5$. Similarly, for the SP75 setup, the injection was done at $\Omega t^{\prime}= 237.5$. We note here that the electric and magnetic fields used for integrating the particles are not a static background and are instead evolving with the fluid as governed by the MHD equations. This enables us to integrate the particle trajectories and energies for a much longer time than what is possible with static background field configurations. This also enables us to capture the effects of transient phenomena such as island contraction and plasmoid movement on the acceleration of particles in a test particle framework. {The results concerning the energetics of the particle runs during the explosive phase are detailed in the paragraphs below.}

\begin{figure}[ht]
    \centering
    \includegraphics[width=0.48\textwidth]{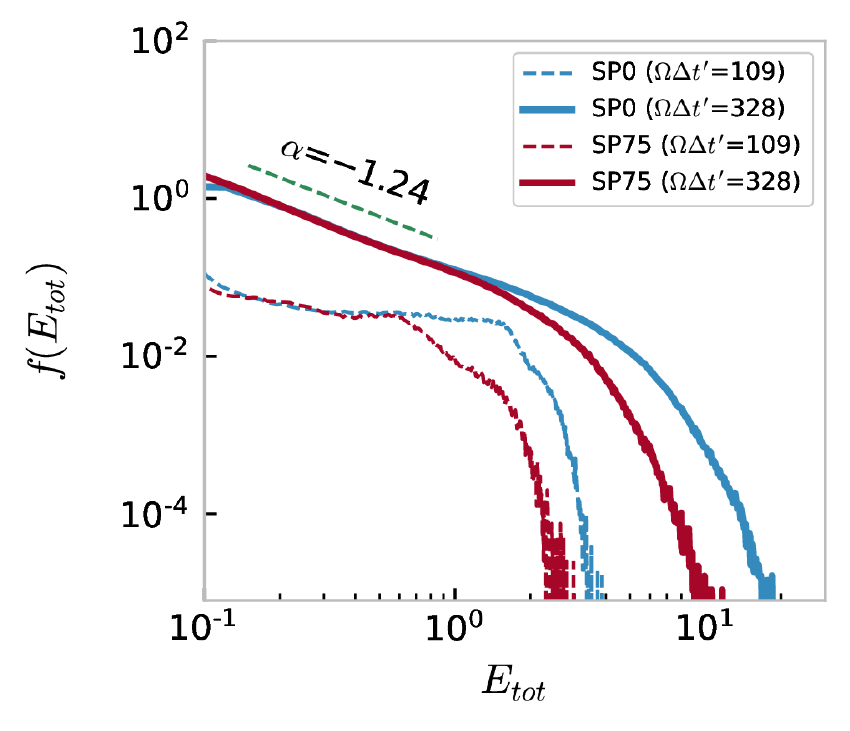}
    \caption{A plot of the normalised energy spectrum of the particles injected during the explosive reconnection phase after an integration time of $\Omega \Delta {t}^{\prime}= 109$ (dashed lines) and $\Omega \Delta {t}^{\prime}= 328$ (solid lines) for the two setups. The blue colored curves correspond to the SP0 setup and the red colored curves correspond to the SP75 setup. The lower limit of the x-axis has been truncated in this plot.}
    \label{fig:histo}
\end{figure}

  {The energy spectrum of the particles were generated by obtaining a normalised energy histogram of all the particles present in the domain at a required time where the quantity ($v^2$/2) was taken as a measure of the particle energy.} For the SP0 setup, after the injection of the particles at $\Omega {t}^{\prime}= 172.5$, a distinct non thermal population giving rise to a higher energy tail compared to the injected distribution quickly begins to show as early as $\Omega { t}^{\prime}= 198$ ($\Omega \Delta {t}^{\prime}= 25.5$). This tail then advances to higher energies showing a tendency towards forming a power-law slope. The dashed curves in Figure \ref{fig:histo} shows the energy spectrum after an integration time of $\Omega \Delta {t}^{\prime}= 109$ ($\sim 1/3$ of the total integration time). As the integration continues, the spectrum eventually gravitates towards a time invariant power-law slope at moderate energies. This time invariant slope can be distinguished as early as $\Omega \Delta {t}^{\prime}= 226$ for both the setups. The power-law has a form of $f(E)= E^{\alpha}$ where $\alpha$ is the spectral index. A statistical likelihood test on the high-energy tail yielded better results for an exponential function fit rather than a power-law. We thus assert that the spectrum of the accelerated particles is a power-law with a high energy exponential cut-off. The final time invariant power-law spectrum attained by the particles after a total integration time of $\Omega \Delta {t}^{\prime}= 328$ can be seen by the solid blue and red curves in Figure \ref{fig:histo} corresponding to SP0 and SP75 setups respectively. It has previously been highlighted in literature that a DC electric field alone cannot produce the double power-law spectrum observed during solar flares\cite{liu_pwrlwexp}. Liu et.al.\cite{liu_pwrlwexp} have highlighted that a set of self consistent but static electric and magnetic fields cannot reproduce the observed double or single power-law spectrum of energetic particles in the DC electric field acceleration scenario. For the ions considered in our study, we employ background electric and magnetic fields that are evolving with time. We even include the effects caused by the Fermi acceleration mechanisms. Even though we obtain a time invariant power-law spectrum for the accelerated particles, we still report an exponential tail at very high energies.  {Hamilton et. al. \cite{Hamilton_2005} have highlighted that the formation of either a power-law or an exponential spectrum formed during particle acceleration at X-type reconnection regions with a finite guide field is sensitive to the choice of the guide field strength and the trapping time of the particles inside the acceleration regions\cite{liu_pwrlwexp, Anastasiadis_1997}. Our study reports a similar combination of a power law with an exponential tail even with the exclusion of a guide field.} The formation of such a spectra however, is a clear indication of non-thermal acceleration. The power-law index was calculated to be $\alpha= {\rm -1.24}$ (shown by the green dashed line above the curves in Figure \ref{fig:histo}).  {Test particle simulations of electrons in an X-point configuration carried out by Liu et.al (2008)\cite{liu_pwrlwexp} have produced a power law index $\alpha= {\rm -1.2}$. Similar test particle runs by Akramov and Baty (2017) for relatively heavier particles (mass= 10 $m_e$) in a static background DTM configuration have also produced a spectral index of $\alpha= {\rm -1.75}$. Werner et. al.(2015) have performed fully kinetic particle-in-cell simulations of magnetic reconnection in pair plasmas consisting of Harris current sheets and have found that the particle energy spectra approaches a value of $\alpha= {\rm -1.75}$ for large system sizes. The spectral index obtained from our test particle simulations evolved in a dynamic background during the explosive reconnection phase in a double current sheet system thus falls within a similar range of what has previously been reported in literature for reconnection in Harris sheets and X-point like structures.} 

The accelerated particle spectrum for the SP75 system is marked by the red curve in Figure \ref{fig:histo}. The time invariant power-law index for this setup shows a negligible difference when compared to the case with no shear. We thus argue that the principal acceleration method in the system responsible for producing the dominant power-law is very similar for both the setups and the presence of a velocity shear does not have any significant effect on this mechanism which results in a negligible effect on the power-law spectral index of the particles. The emergence of such a power-law spectrum with a similar spectral index is an outcome of the mechanism responsible for the acceleration of the particles\cite{guo2015,guo_2019} and for the setups SP0 and SP75, the acceleration mechanism broadly remains the same with the principal difference lying in the efficiency of the mechanism as will be evident from section \ref{prtcl_traj}. The difference thus occurs in the high energy tail of the spectrum which truncates at a lower energy than the SP0 setup. Further investigations into the reason for this difference has been described in the following sections.

\subsection{\label{prtcl_traj}Particle Trajectory}

We select and analyze a few of the particle trajectories in the domain to accurately determine the position where the acceleration occurs and the amount of acceleration caused by various components in the domain such as moving structures, turbulence, and direct electric field during the explosive phase. For a better representation of the energetics, the total kinetic energy per unit mass ($E_{\rm tot} = v^2/2$) of the particle was separated into two components, the in plane component ($E_{IP}$) representing the acceleration in the x-y plane and the out-of-plane component ($E_{OP}$) representing the acceleration along the z-direction.

\begin{figure*}[ht]
    \centering
    \includegraphics[width=2.0\columnwidth]{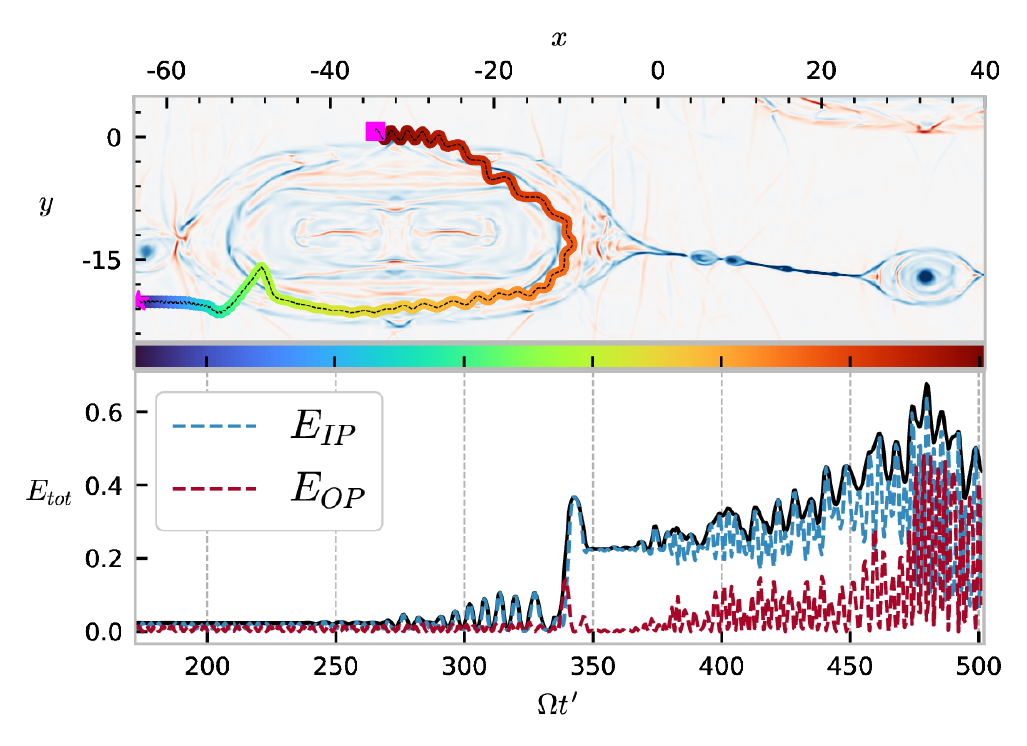}
    \caption{Trajectory of a particle that is slightly accelerated from the SP0 setup. The magenta star and the square represent the start and the end of the trajectory respectively. The background in the top panel shows a snapshot of the current density (\textbf{J}) at $\Omega {t}^{\prime}= 428$ which is almost at the mid point of the energizing process. The color-bar between the panels correspond to the time given in the abscissa of the bottom panel. The ratio of the final kinetic energy attained by the particle to its initial energy is $R_{\rm KE}$=18.}
    \label{fig:traj_low}
\end{figure*}

We, first look at some particles that are slightly accelerated in the domain. One such particle is shown in Figure \ref{fig:traj_low}. The particle initially is wandering close to the largest magnetic island in a direction as determined by its initial velocity components. It then starts to drift slowly towards the current sheet where it abruptly gets hit by the outer boundary of a moving plasmoid. This flings the particle in the downward (-y) direction (at $x\sim -48$) which causes a spike in the {in-plane} component of the kinetic energy of the particle at about $\Omega {t}^{\prime} \sim 345$ as seen in the bottom panel of Figure \ref{fig:traj_low}. This impulsive mechanism is responsible for about half of the total energy gained by the particle in a very small amount of time. The particle then travels along the outer boundary of the monster-plasmoid and slowly keeps energizing due to the magnetic inhomogeneities present inside the island which further increases the {in-plane} energy of the particle. {Generally, there can be multiple mechanisms by which the particles gain energy from magnetic inhomogeneities. Firstly, the turbulent magnetic structures can provide scattering centers for the particle to energize by Fermi processes\cite{Fermi_1949}, secondly, a non uniform magnetic field encountered by the particle during each gyro-rotation may lead to an overall work done on the particle by the Lorentz force\cite{Kowal_2011}. Even though it is difficult to isolate the dominant mechanism in this case, the change of the pitch angle of the particle during its motion indicate that the Fermi acceleration mechanism might be the effective mechanism in this case.} Towards the end of the trajectory, the particle also encounters a thin region of enhanced electric fields that increases the energy of the particle in the out-of-plane direction as it performs multiple passes through the region due to its gyrating motion. However, most of the overall kinetic energy of the particle is provided by the {in-plane} acceleration. The kinetic energy per unit mass ($E_{\rm tot}$) of this slightly accelerated particle reaches a maximum value of 0.67$v_{\rm A}^2$ throughout the run. We define the ratio of the final value of $E_{\rm tot}$ attained by the particle to its initial value by a parameter named as $R_{\rm KE}$. For this slightly accelerated particle, a value of $R_{\rm KE}$=18 was calculated. This shows that for the slightly accelerated particles in the domain, most of the energy is imparted by the scattering of the particles with the moving structures (such as plasmoids) in the domain.

\begin{figure*}[ht]
    \centering
    \includegraphics[width=2.0\columnwidth]{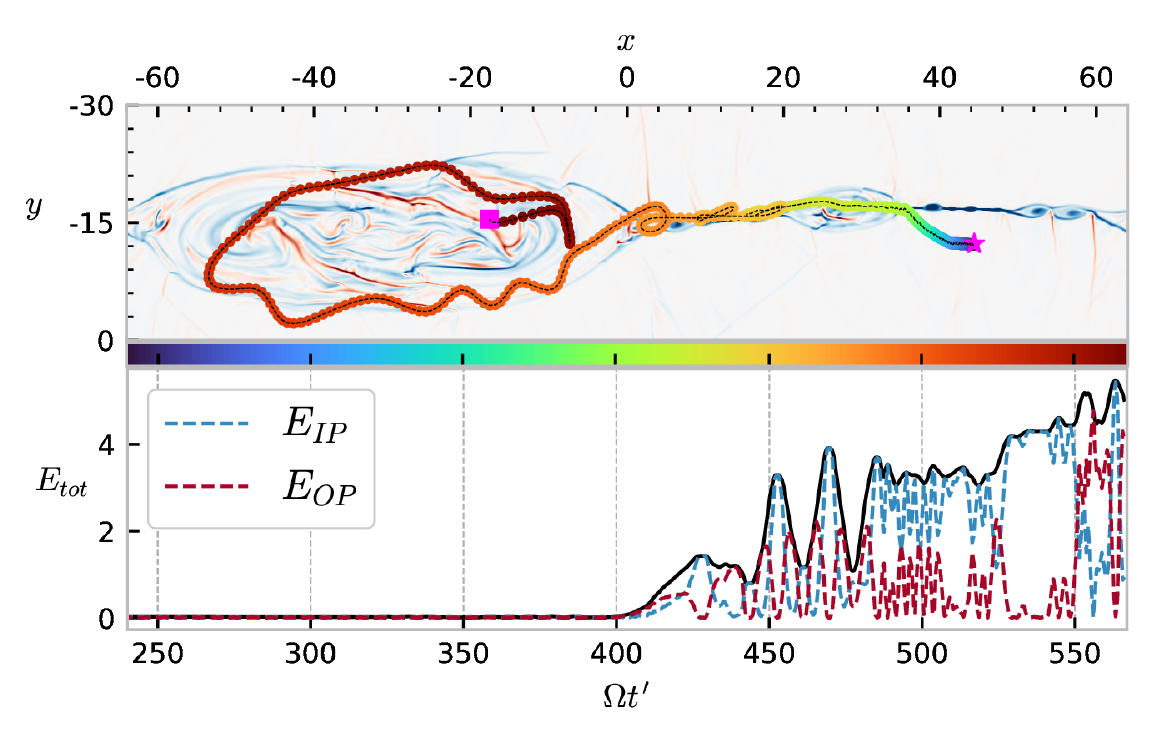}
    \caption{Trajectory of the particle that is moderately accelerated from the SP75 setup with a shear speed of $0.75v_{\rm A}$. The magenta star and square represent the start and the end of the trajectory respectively. The background in the top panel shows a snapshot of the current density at $\Omega {t}^{\prime}= 493$ which is almost at the mid point of the energizing process. The color-bar between the panels correspond to the time given in the abscissa of the bottom panel. The ratio of the final kinetic energy attained by the particle to its initial energy is $R_{\rm KE}$=214.}
    \label{fig:traj_med_vsh_75}
\end{figure*}

Irrespective of the presence of a shear, the trajectories of almost all the moderately accelerated particles follow a common pattern for both the shear speeds. They initially drift into the current layer where plasmoids are forming, there,  they either get injected into one of the small plasmoids that are being ejected and get advected along the current sheets or, are `picked up' by the outflow exhausts and injected into the monster-plasmoids. This pickup mechanism has been previously observed in simulations and is presumed to be similar to the mechanism of pickup of a newly ionised particle in the solar wind\cite{Drake_2009, Mobius_1985, Eriksson_2020}. Thereafter, they are expelled from the smaller plasmoid into either of the monster plasmoids where the guiding center of the particle travels in an orbit traversing the boundary of the island. This is exactly what is seen from Figure \ref{fig:traj_med_vsh_75} which shows a particle from the SP75 setup. The particle initially drifts into the current sheet and is injected into a small plasmoid that has just formed and is being advected along the current sheet. Throughout the time that the particle is moving along with the small plasmoid, it also bounces around inside while reflecting off of its boundaries (as seen from the small loops in the trajectory in the region $x$ = 0 to 20). This reflecting mechanism leads to a considerable increase in the {in-plane} component of the energy whereas there is a spike in the out-of-plane component wherever the particle encounters the convective electric field produced by the small scale exhausts inside the plasmoids. This combined effect results in an increase of the {in-plane} as well as the out-of-plane component of the energy of the particle. After the particle enters the big magnetic island and gyrates along the boundary, its energy is further increased as it interacts with the turbulent fluid inside the island. This particle finally attains a value of $R_{\rm KE}$=214 by the end of the acceleration process. We also see that such moderately accelerated particles are reflected at a certain angle to their direction of approach from the boundaries of the monster plasmoids as seen in Figure \ref{fig:traj_med_vsh_75}.

\begin{figure*}[ht]
    \centering
    \includegraphics[width=2.0\columnwidth]{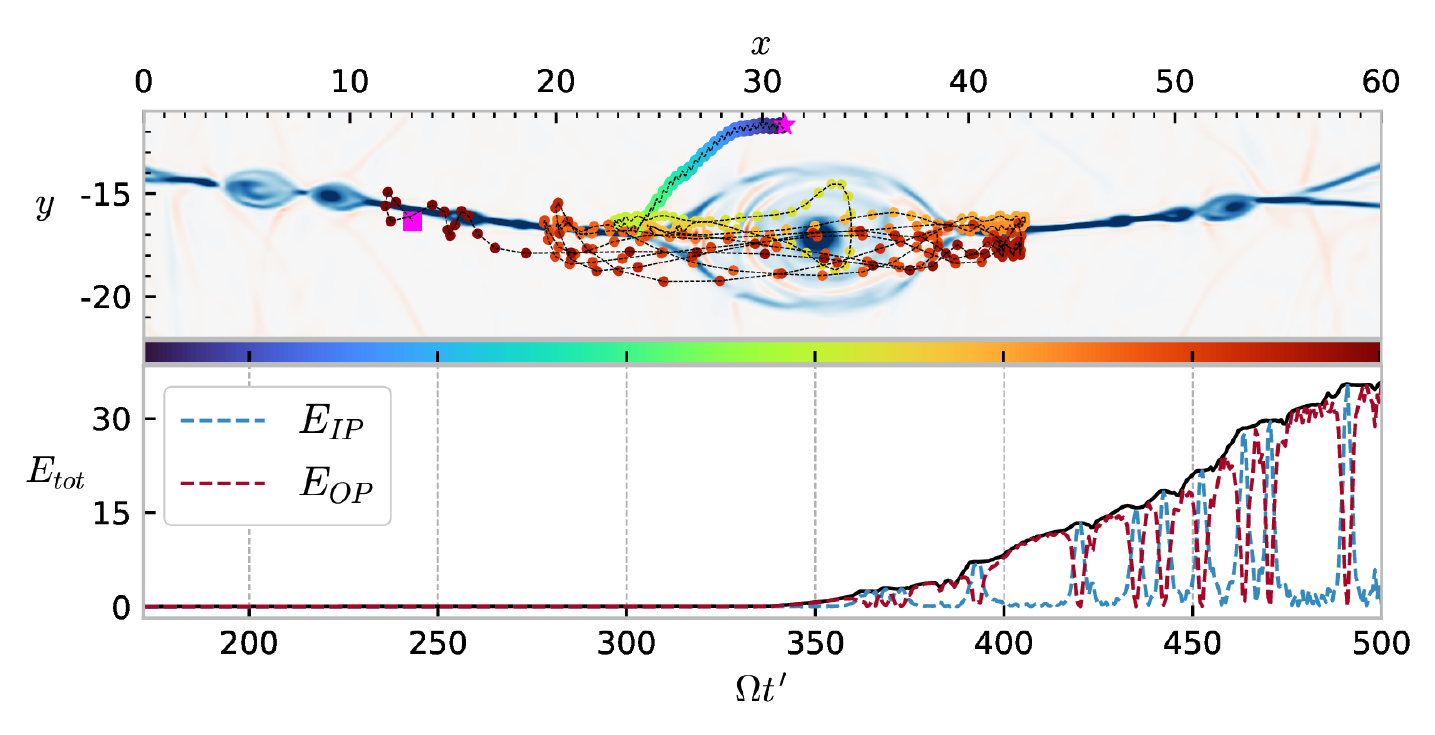}
    \caption{Trajectory of the particle with the highest energy from the SP0 setup. Top panel shows a small portion of the whole domain. The magenta star and the magenta square represent the start and the end of the trajectory respectively.  The background in the top panel shows a snapshot of the current density at $\Omega {t}^{\prime}= 428$ which is almost at the mid point of the energizing process. The color-bar between the panels correspond to the time given in the abscissa of the bottom panel. The ratio of the final kinetic energy attained by the particle to its initial energy is $R_{\rm KE}$=1677.}
    \label{fig:traj_highest}
\end{figure*}

For the particle run with zero shear (SP0), interestingly, the particles with the highest energy in the domain were the ones trapped in the segment between the two sides of the largest magnetic island of either of the current sheets, i.e in the region of where plasmoids were continuously being produced and ejected from multiple X-points during the non-linear reconnection phase. The top panel of figure \ref{fig:traj_highest} shows the trajectory of one such particle. The bottom panel of Figure \ref{fig:traj_highest} shows the time evolution of the energy of the particle. 

The particle, in Figure \ref{fig:traj_highest} initially drifts in from the upper side of the lower current sheet, and encounters the current layer at $\Omega {t}^{\prime} \sim 335$. This particle is then pulled into a small plasmoid where it is reflected back and forth multiple times by the coalescence of smaller plasmoids from both sides. This particle trapped in the plasmoid unstable region finally attains a value of $R_{\rm KE}$=1677. The strong electric fields present in this region is the dominant cause of the acceleration of such particles which is clear from the energy-time plot in Figure \ref{fig:traj_highest}. The majority of the increase in the total energy is provided by the out-of-plane component of the energy of the particle, however, the collision of the particles with the small scale reconnection outflows (that increases $E_{IP}$) inside the small plasmoids also impart a small fraction of the total energy.

Another class of high energy particles show a slightly different acceleration mechanism which is a combination of the electric fields, energization caused by shrinking magnetic islands, and other dynamic effects such as the {scattering of the particles from the reconnection exhaust-fronts inside the monster plasmoids.}

\begin{figure*}[ht]
    \centering
    \includegraphics[width=2.0\columnwidth]{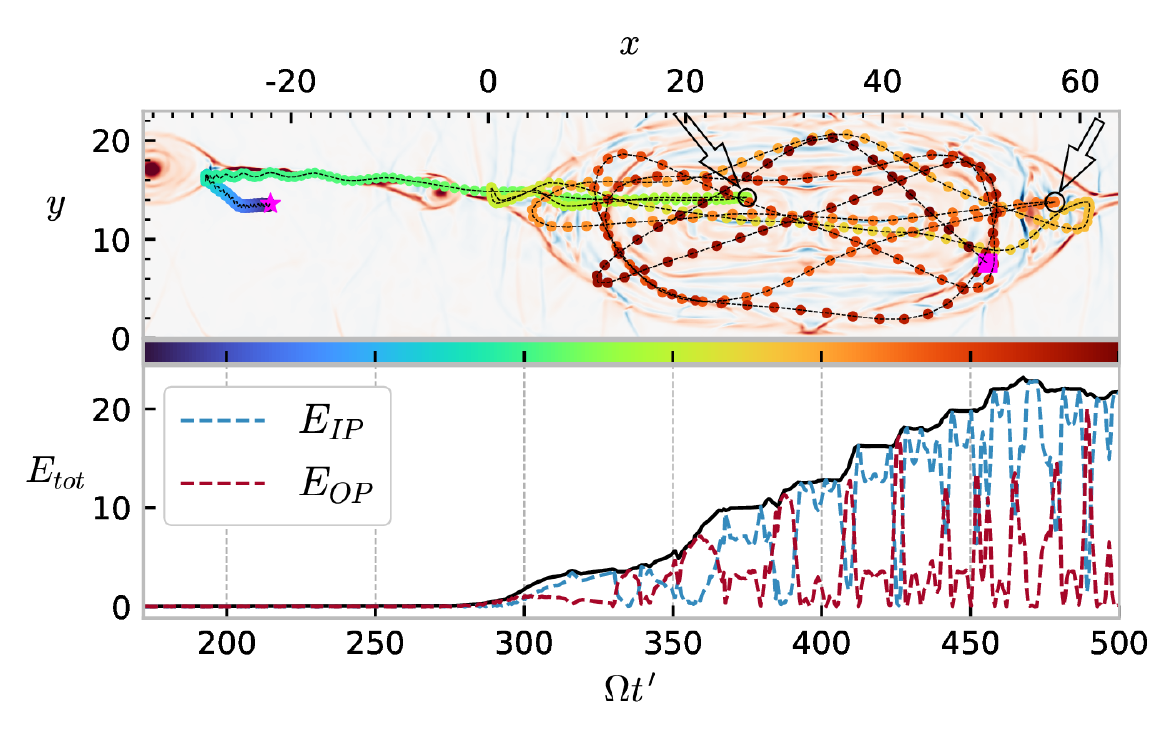}
    \caption{Trajectory of a particle from the SP0 setup having a rather different acceleration mechanism governed by the island dynamics. The background in the top panel shows a snapshot of the current density at $\Omega {t}^{\prime}= 428$ which is almost at the mid point of the energizing process. The arrows point towards the encircled regions where the particles encounter a nearly head-on collision. The color-bar between the panels correspond to the time given in the abscissa of the bottom panel. The ratio of the final kinetic energy attained by the particle to its initial energy is $R_{\rm KE}$=567.}
    \label{fig:traj_high}
\end{figure*}

\begin{figure}[ht]
    \centering
    \includegraphics[width=\columnwidth]{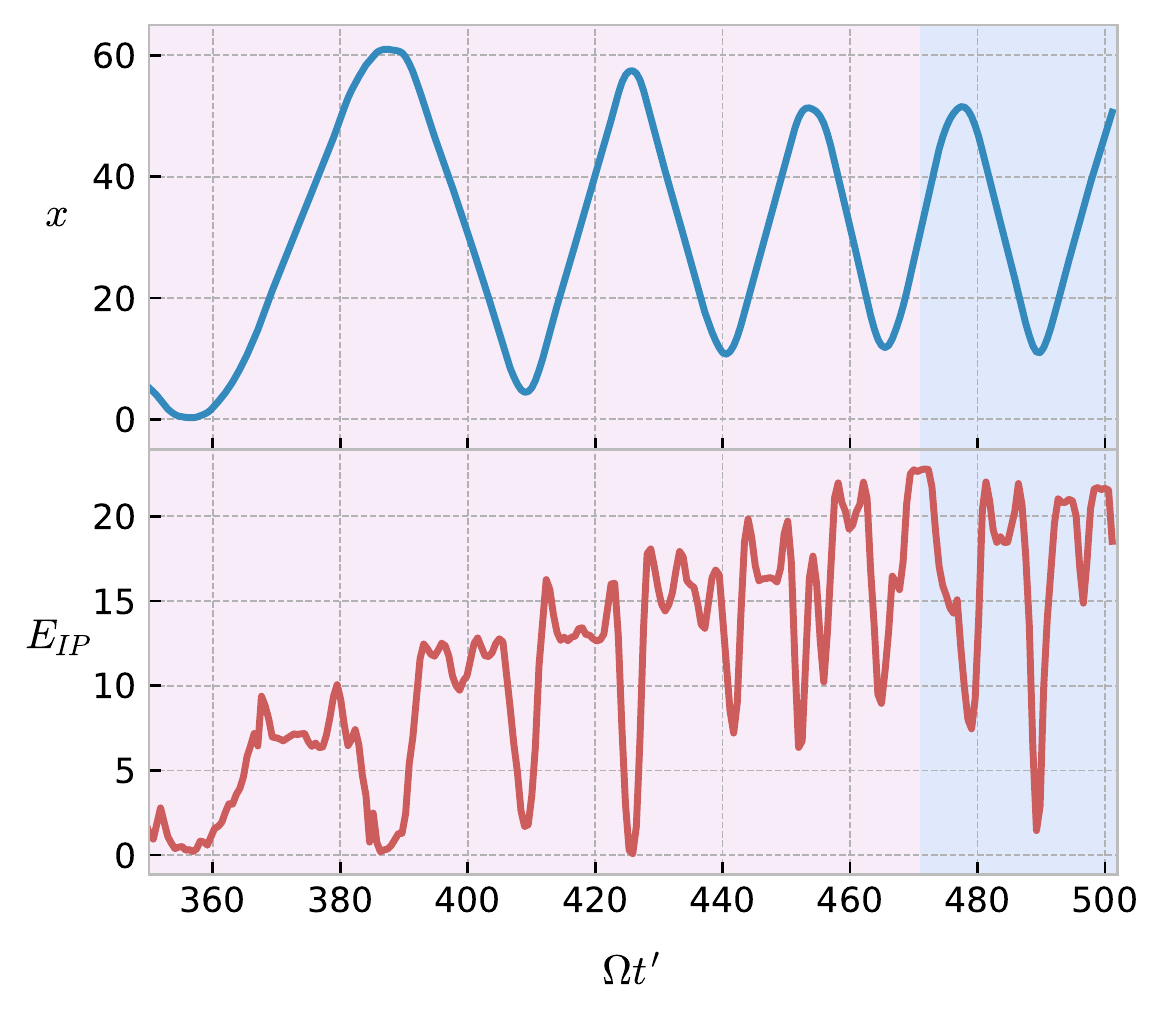}
    \caption{Top panel of the figure plots the x-coordinate of the particle in Figure \ref{fig:traj_high} with time after it enters one of the large magnetic islands. The oscillatory pattern with a decaying amplitude is a result of the particle getting {reflected} back by the converging outflow exhausts from both sides of the islands. The two colors in the plot background stand for two different time stages. The reddish hue represents the time when the span of the x-coordinate is shrinking indicating an increase in energy as can be seen in the corresponding bottom panel. The bluish hue represents the time when the shrinking subsides and a corresponding stabilization of the {in-plane} energy ($E_{IP}$) can be seen in the panel below.}
    \label{fig:conv_flows}
\end{figure}

The particle shown in Figure \ref{fig:traj_high}  also drifts into the current sheet initially like the others. This drift is mainly due to the increase in gradient of the magnetic fields as the particle streams closer to the current sheet. Once it reaches the current sheet, it is given a significant amount of {in-plane} acceleration as it is advected by the plasmoids moving out of the X-points and is pushed into one of the largest magnetic islands after the transporter plasmoid merges with it. The first reflection of the particle is by the turbulence caused by the interaction of the reconnection exhausts during the beginning of the explosive phase from both sides at the center of the monster plasmoid. After that, the particle gets {reflected} back and forth by the reconnection outflow jets from both edges of the monster-plasmoid. This repetitive scattering effect causes a significant increase in the energy of the particle which is essentially a first order Fermi acceleration\cite{particle_acc_in_mag_recon_dalpino}. Fermi acceleration mechanisms have indeed been established as the dominant mechanism of acceleration in reconnection environments\cite{guo_2019}. We note that the particles that scatter nearly head-on from the reconnection exhausts at the two ends of the monster plasmoids (observed here in the case of high energy particles, e.g the particle shown in Figure \ref{fig:traj_high} encounters nearly head-on collisions at the encircled regions marked by arrows) gain more $E_{IP}$ than other particles that suffer collisions at a certain angle of approach, this is an expected result as such collisions lead to a more efficient energy transfer. The collision of the particles with the reconnection exhausts provide a two-fold acceleration mechanism. The first mechanism is that the collision leads to a {reflection} of the particle at a larger speed compared to its initial approach speed. This causes an increase in the {in-plane} component of the energy.  The other is the out-of-plane acceleration caused mainly due to the strong $-\mathbf{v} \times \mathbf{B}$ electric field present in the region where the reconnection outflow enters the monster-plasmoid. {Even though these out of plane electric fields of smaller magnitudes are present in other regions along the exhausts from the X-points, the particle transits through such regions fairly quickly. In contrast it spends a significant amount of time inside the monster plasmoids where it suffers multiple collisions with the reconnection exhaust fronts, this scattering mechanism is thus more effective in the overall acceleration of the particle.} The acceleration caused by the reflection of the particles from the converging reconnection exhausts inside the large magnetic island can be seen in Figure \ref{fig:conv_flows}. It is seen that in between $\Omega {t}^{\prime} \sim 360$ and $\Omega {t}^{\prime} \sim 470$ the particle is repeatedly {reflected} back and forth as it is slammed by the counter streaming exhausts moving towards each other inside the large magnetic island which reduces the X-coordinate span of the particle. This phase is marked in red hue in Figure \ref{fig:conv_flows}. The {in-plane} component of the energy shows an increasing trend during this period. This repetitive mechanism is responsible for an overall increase in $E_{IP}$ by more than 20 times (see Figure \ref{fig:conv_flows}) which results in an increase of the total energy ($E_{\rm tot}$) of the particle shown in Figure \ref{fig:traj_high} to more than twice the value before the onset of this mechanism.

To probe into the difference between the particle spectrum for the S75 setup, we further investigate the particle trajectories for the high energy particles in terms of the maximum energy reached by them during the time integration. We only look into the high energy particles as the difference between the spectra of the SP0 and the SP75 systems is only towards the high energy tail ($E_{\rm tot} \geq 2.0$) as seen from Figure \ref{fig:histo}. A closer look into the trajectories of the slightly accelerated particles show no significant difference for the two setups. An analysis of such particles show that the acceleration mechanism as well as the energy gain for such slightly accelerated particles are exactly the same with and without the presence of shear. They do not encounter regions of very strong electric fields and mainly accelerate while advancing with the fast moving small plasmoids which imparts a significant amount of the initial acceleration. This predominantly increases the {in-plane} component of the energy ($E_{IP}$). {Thereafter, most of the energy is provided by the motion of the particle through magnetic inhomogeneities and turbulence present inside the monster plasmoids as the turbulence provides multiple scattering centers for the occurrence of a stochastic acceleration process.} For the moderately accelerated particles, they initially get injected into the plasmoid unstable region where they are given a significant {in-plane} acceleration and are thereby injected into either of the monster plasmoids. The moderate energy particles are also found to never collide head-on with the reconnection exhausts inside the monster plasmoid.

It is obvious from the energy spectrum in Figure \ref{fig:histo} that the maximum energies attained by the highest energy particles are lesser for the case with a significant shear (SP75). The principal reason for this being that the presence of a shear flow suppresses the efficiency of plasma outflow from the X-points and hence results in a corresponding suppression of the reconnection rate. This also essentially means that the reconnection exhausts are moving slower in the presence of a significant shear flow\cite{cassak_2011}. 

\begin{figure}[ht]
    \centering
    \includegraphics[width=\columnwidth]{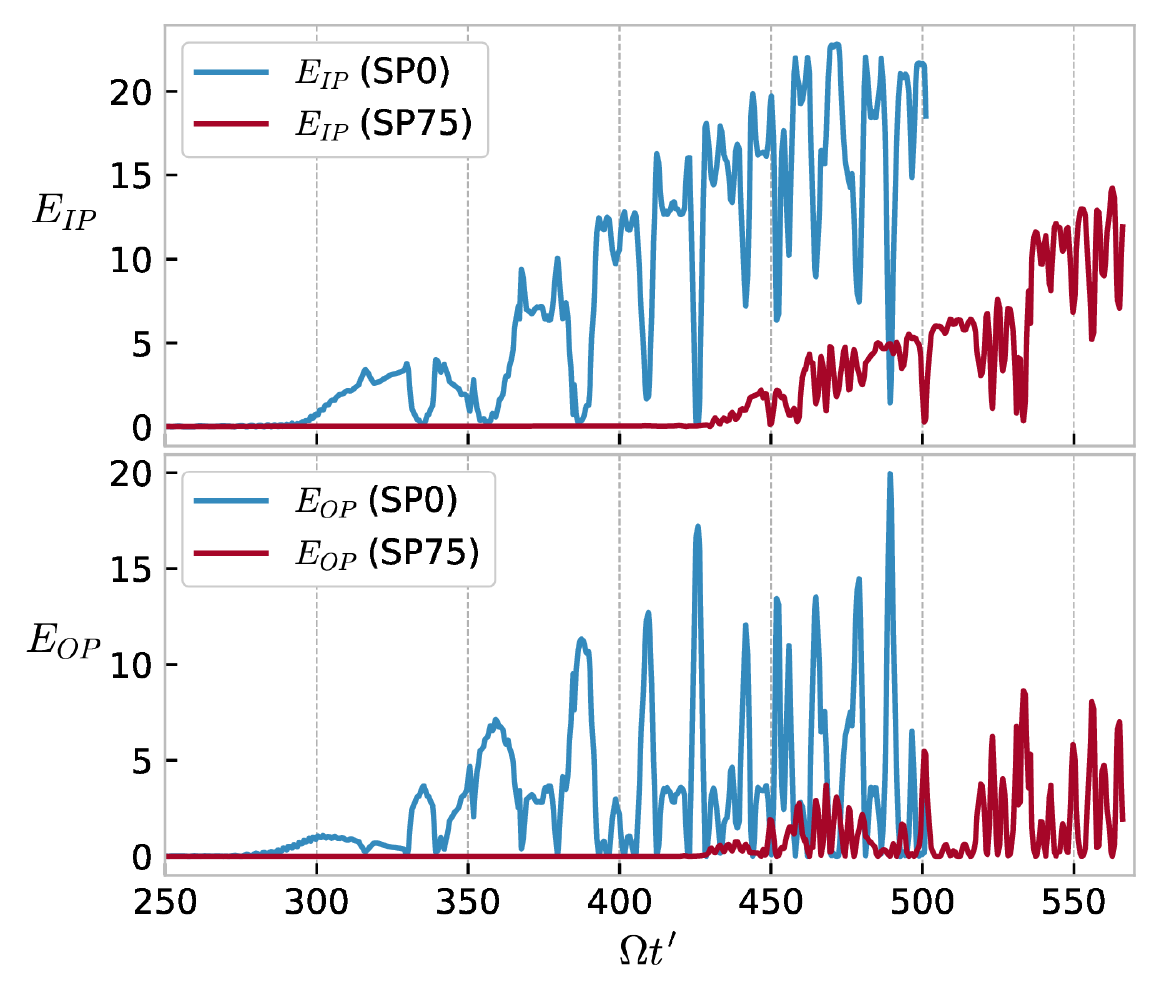}
    \caption{A comparison of the {in-plane} and and out-of-plane energization of two high energy particles having a very similar trajectory, one from the SP0 setup and another from the SP75 setup.}
    \label{fig:par_perp}
\end{figure}

Thus, the reconnection exhausts in the presence of a significant shear speed (in SP75) are slower compared to the setup with zero shear (SP0). This implies that the particles {reflect} back slower after colliding with the exhausts in the SP75 setup which essentially means that the {in-plane} energy imparted per collision is lesser in the presence of a shear flow. This can be readily seen in the top panel of  Figure \ref{fig:par_perp} which shows that the increase in the {in-plane} component of the kinetic energy is significantly faster for the SP0 setup when compared to the SP75 setup.

{Even though a $B_{y}$ component of the magnetic field exists all along the reconnection outflow region, the fast outflow exhausts, when they enter the monster plasmoids, interacts with the $B_y$ component of the predominantly O-type magnetic field structures present inside the plasmoids. This gives rise to a pile-up of the $B_y$ component which causes a significant increase in the convective electric fields ($-\mathbf{v_x} \times \mathbf{B_y}$) in the region where the exhausts enter the monster plasmoids.} {These are also the regions which contain the reconnection exhaust fronts that the particles reflect from}, and thus we see a spike in the out-of-plane component of the particle kinetic energy ($E_{OP}$) wherever there is a {reflection} from such regions. As a comparison, we have considered two different particles, one from each setup (SP0 and SP75) showing a similar trajectory pattern. We have then calculated the average value of the magnitude of the electric field produced in the exhaust region ($51.0 \leq x \leq 68.0$ and $8.0 \leq y \leq 20.0$) for the two cases during a time snap when the two particles were interacting with the reconnection exhausts ($\Omega {t}^{\prime} = 426$ for SP0 and $\Omega {t}^{\prime} = 533$ for SP75). The results show that the SP0 setup has an average electric field that is more than twice the SP75 setup.  Therefore, the acceleration of the particle in the out-of-plane direction is significantly higher in the absence of a shear (SP0 setup). Thus the two mechanisms, {(a) particles reflecting back at an increased in-plane speed after scattering from the reconnection exhausts,} (b) acceleration by stronger convective electric fields in the outflow regions, combined together result in a more efficient acceleration in the absence of a shear flow in the SP0 setup as compared to the SP75 setup.  

\section{\label{Summary}Discussions and Summary\protect}

We have investigated the time evolution of a double current sheet system within the resistive magnetohydrodynamic framework with special emphasis on the explosive reconnection phase. Particle acceleration during the explosive phase has also been studied with the help of test particles (protons). We evolve the particles in a moving fluid background using the MHD-PIC framework to capture the effects of transient phenomena developing on a fast timescale on particle acceleration. The key points of this study are highlighted below.

\begin{itemize}
    \item A study of the fluid evolution shows three distinct phases, an initial slow growth phase, an explosive non-linear growth phase and a final relaxation phase. These three stages can be clearly distinguished from the evolution of the change in the kinetic energy density in Figure \ref{fig:magen_kinen_sh0} for the S0 system. In the absence of a shear, the initial transient phase lasts for $\Delta t/\tau_{\rm A} = 175$. During this phase the primary magnetic islands start to grow gradually and the reconnection proceeds in a Sweet-Parker fashion. The period that follows the transient slow growth phase is an explosive non-linear phase where a portion of the current sheet breaks down into numerous small plasmoids. This marks the onset of the plasmoid instability which gives rise to multiple X-points that greatly enhance the reconnection rate giving rise to a fast explosive decrease in the magnetic energy density of the system and a corresponding increase in the overall kinetic energy density of the system. {An interesting difference between the evolution of our setup and DTMs that has previously been studied in literature (e.g Akramov, Baty (2017) \cite{akramov_baty}, Nemati et. al. (2016) \cite{nemati_2016}) is that the plasmoid instability is our setups start by the virtue of a very high value of the Lundquist number of our system, whereas, previous studies dealt with the plasmoid formation at a much later phase of evolution when one of the current sheets are pinched by the monster plasmoids associated with the other.} {The interaction between the large primary magnetic islands and the current sheets bends the layers causing the outflow exhausts entering from both sides inside the small plasmoids to be slightly displaced with respect to each other along the y-direction leading to the formation of fluid vortices inside such plasmoids.} These vortices form even in the absence of a shear flow and could serve as important factors in imparting a twist in the magnetic field lines as has been seen in the 3D helical structure of magnetic field lines inside flux ropes. During the later stages of the evolution, the two large magnetic islands occupy a major portion along the length of the initial current sheets which reduces the effective length of the plasmoid unstable region. This essentially reduces the reconnection rate as can be seen in the flattening of the $\Delta E_{\rm mag}$ profile in Figure \ref{fig:magen_kinen_sh0}.
    
    \item All the other setups where a sub-Alfv\'enic shear speed is present, show a similar three phase pattern which is shifted in time that indicates that the presence of a shear does not alter the overall dynamic behavior of the system, albeit rendering the whole evolution delayed in time by a certain amount. This can also be confirmed by the convergence of the final states and energies in all the systems. This delay is caused by the suppression of the instability growth during the linear and early non-linear phase of the plasmoid instability which in turn delays the transition to the fast explosive reconnection phase. We have also verified that no fast-reconnection dynamics were observed for the case with a super-Alfv\'enic shear flow. 
    
    \item A study of the reconnection rate during the very early phases of the system evolution shows results that are consistent with the theoretical scaling given by equation \ref{eqn:recon_shear}. The reconnection rate was measured as the average of the magnitude of the electric field in two thin strips along the current sheet lengths. A slight deviation is observed for the case of S100. To verify the validity of the theoretical scaling for the explosive phase of the coupled tearing mode, a further analysis of the reconnection rate was performed at a later stage after the onset of the plasmoid instability. {Our results show that the scaling fails for higher shear speeds (S50 and S75) if one ignores the dependence of the reconnection rate on the size of the magnetic islands (islands having larger widths tend to enhance the reconnection rate). This can be seen by the deviation of the blue hexagons on the bottom panel of Figure \ref{fig:rec_rate_vs_shear} from the theoretical curve when the reconnection rate was measured for all the systems during the same time span.} The width of the magnetic islands for the S0 and S25 setups are higher which results in a stronger coupling between the current sheet layers that enhances the reconnection rate for these systems. We have also measured the reconnection rate for S50 and S75 setups at later times when the island half widths for these setups were equal to that of the setups S0 and S25 when the initial measurement of the reconnection rates were done. The same is plotted in red triangles on the bottom panel of Figure \ref{fig:rec_rate_vs_shear} which shows an excellent agreement with the theoretical scaling. We thus infer that the theoretical scaling of the reconnection rate for a single current sheet holds true even for the explosive phase of the DTM like coupled tearing mode if the feedback effects arising due to the structure of the magnetic islands are taken into account.
    
    \item Particle acceleration during the explosive phase of the evolution was also studied using test particles in a dynamic fluid background for the setups SP0 and SP75. {The mechanisms can be condensed into two main processes (a) acceleration by the electric fields (b)scattering from dynamic structures (i.e plasmoids, reconnection exhaust fronts) in the domain.} A large role in the acceleration of the particles were played by the convective electric fields and the interaction of the particles with the reconnection exhausts. Previous studies have found that X-line acceleration may not be significant in the overall energization of the particles\cite{guo_2019}. Our results are consistent with this result as for all the high energy particles in our domain the Fermi acceleration mechanisms are the dominant process for the acceleration of high-energy particles. These Fermi mechanisms are preceded by the interaction of the particles with an X-point and the associated reconnection exhausts that imparts only a small amount of energy to the particle and helps in its injection into the plasmoids. However, this mechanism  is important for other acceleration processes that might follow, as it is a well known fact that the particles need to have speeds comparable to or greater than the Alfv\'en speed to be efficiently accelerated by the processes such as island contraction and {first order Fermi mechanism} during interaction of particles with reconnection exhausts inside the plasmoids\cite{Drake_2009}. For all the high energy particles studied in our simulations, the pickup mechanism and the acceleration at the X-points provides enough energy to the particle to render it super-Alfv\'enic.
    
    \item The accelerated particles in both the SP0 and SP75 setups form a time invariant power-law spectrum with an exponential tail at high energies. The power-law index is identical for both the cases which confirms that the presence of a velocity shear has negligible effect in determining the power-law index of the spectrum. The difference is noticeable however, in the exponential tail wherein, the very high energy particles in the SP75 setup had lesser kinetic energy than the SP0 setup by a factor of about two. We attribute this to two reasons. Firstly, the presence of a shear slows down the speed of the outflow exhausts from the reconnection regions\cite{cassak_2011}. A principal mechanism of the acceleration of high energy particles in the domain is due to their {reflection from the outflow exhausts} entering both sides of the monster-plasmoids. Thus, the slower exhausts from the SP75 setup results in a lesser {in-plane} acceleration of the particles as is evident from the top panel of  Figure \ref{fig:par_perp}. {Secondly, the exhausts entering the monster plasmoids result in a pile up of the $B_y$ component of the magnetic field along the plasmoid edges enhancing its strength in those regions. This strong $B_y$ component further enhances the convective electric fields which accelerate the particles in the out-of-plane direction. The higher value of the reconnection exhaust speeds thus result in a stronger electric field for the SP0 setup (nearly double that of SP75) which leads to a stronger out-of-plane acceleration for the particles in SP0.} These two mechanisms together contribute to the difference in the high energy tail region of the particle spectra.
    
    \item The maximum speed reached by the particles was nearly of the order of 0.1C where `C' is the artificial speed of light set to be 100$v_{\rm A}$ in the code. It is thus obvious that such particles are having supra-thermal energies and can easily be produced just by the processes associated with magnetic reconnection. It is known that a large number of supra-thermal `seed-particles' are required for shock-acceleration processes to be efficient\cite{Provornikova_2016} and therefore such a population of supra-thermal ions, as obtained from our simulations, can in fact play the role of said seed particles. It has also been highlighted that more than 92\% of all SEP events are accompanied by either type-II or type-III radio bursts\cite{Winter_2015}. These radio signatures indicate the presence of supra-thermal particles in the flaring regions and hence the notion that such particles can form the initial seed population that can further be accelerated by supplementary mechanisms is compelling. The acceleration received by the particles in our framework is directly proportional to the value of $\alpha_{\rm p}$. For lighter particles (say electrons) the amount of acceleration received in each time-step would thus be about three orders of magnitude higher which would possibly result in their acceleration to relativistic speeds, however, the computational cost to resolve the gyration scale of such lighter particles in a moving fluid background would be profuse. The precise effects of such electron scale physics is more consistently addressed by either a fully kinetic PIC scheme\cite{Wilson_2016, Doss_2016, Travnicek_2009, Shimizu_2016} or via coupling the implicit PIC scheme with MHD fluid codes\cite{Maknawa_2018, Chen_2017,sugiyama_2006}. This is currently outside the scope of our study.   
\end{itemize}

  In essence, our study includes the dynamics of a double current sheet system in the presence of a shear flow across the current sheets, we have also demonstrated the consequence of such DTM like dynamics on the process of particle acceleration and probed into the exact mechanisms by which ions are energized. 
  In this study, we have not considered the presence of a guide field which can affect particle acceleration due to ion heating \cite{Drake_2009}. Additionally, including asymmetries in density and magnetic fields along with feedback effects of the CR-particles to the underlying fluid via coupling with implicit PIC scheme shall be addressed in a future study.  

\section*{Acknowledgement}

The authors thank the the reviewers for their constructive comments and useful suggestions that have significantly improved the manuscript. AP is a research scholar at IIT Indore and is thankful for the support provided by the institute. AP also thanks Sayan Kundu, Sriyasriti Acharya, Gourab Giri and Indu Kalpa Dihingia for the fruitful discussion sessions during the course of the work. BV would like to acknowledge the support from the Max Planck partner Group Award. All computations presented in this work have been carried out using the facilities provided at IIT Indore and the Max Planck Institute for Astronomy Cluster: ISAAC which is a part of the Max Planck Computing and data Facility (MPCDF).

\section*{Data Availability Statement}
The data that support the findings of this study are available from the corresponding author upon reasonable request.

\typeout{}

\bibliographystyle{aipauth4-1}
\bibliography{ms_ap}

\newpage

\appendix*

\section{Resolution Study}
With regard to analyzing the accuracy of the PLUTO code to correctly capture all the transient effects, we have performed a resolution study to properly identify the sufficient resolution required to capture the small scale phenomena such as the plasmoid instability caused by the thin current sheets. Firstly, we emphasise here that the solver we are using is HLLD, which is the most accurate among the HLL-type solvers that exclude the eigen-decomposition. In conjunction with HLDD, the Vanleer flux limiter ensures less numerical diffusion in comparison to other limiters such as minmod. We perform simulations of various resolutions for the setup with no shear to identify the the resolution where a modest convergence is achieved. The resolutions employed were $320\times480$, $640\times960$, $1280\times1920$, $1920\times2880$, $2560\times3840$ corresponding to grid sizes ($\Delta x$ or $\Delta y$) of 0.4, 0.2, 0.1, 0.066, 0.05 respectively. We note here that initially the current sheet has a thickness of $\Delta x= 1.0$. This thickness gradually decreases due to the thinning with increasing time. Analysis of the same in section \ref{Results:General_evolution} show that the disruption of the current sheets into the plasmoid instability occur when the thickness (measured as the FWHM of the current density, \textbf{J}) of the current sheets approach $\Delta x \sim 0.37$. This is also in accordance with the theory of current sheet disruption given by Pucci and Velli ; 2014 (see text for details)\cite{Pucci_2013}.

\begin{figure}[ht]
    \centering
    \includegraphics[width=\columnwidth]{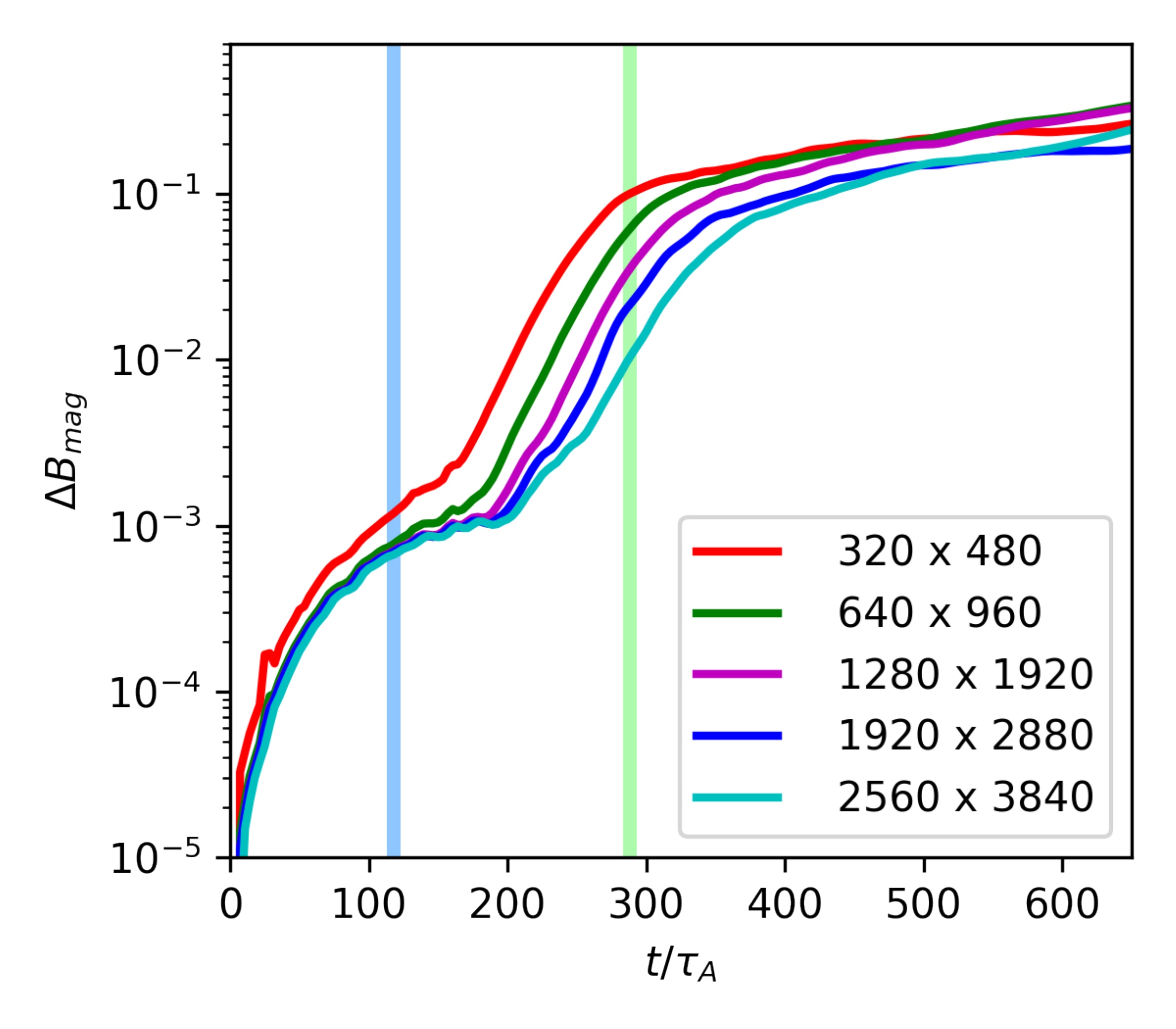}
    \caption{Time evolution of the magnetic energy density ($\Delta E_{\rm mag} = E_{\rm mag(t=0)} - E_{\rm mag(t)}$) of of S0 for various resolutions. The blue and the green vertical strips correspond to the duration when the reconnection rates for section \ref{Sec:Recon_Rate} were measured.}
    \label{fig:res_test}
\end{figure}\begin{figure*}[ht]
    \centering
    \includegraphics[width=\textwidth]{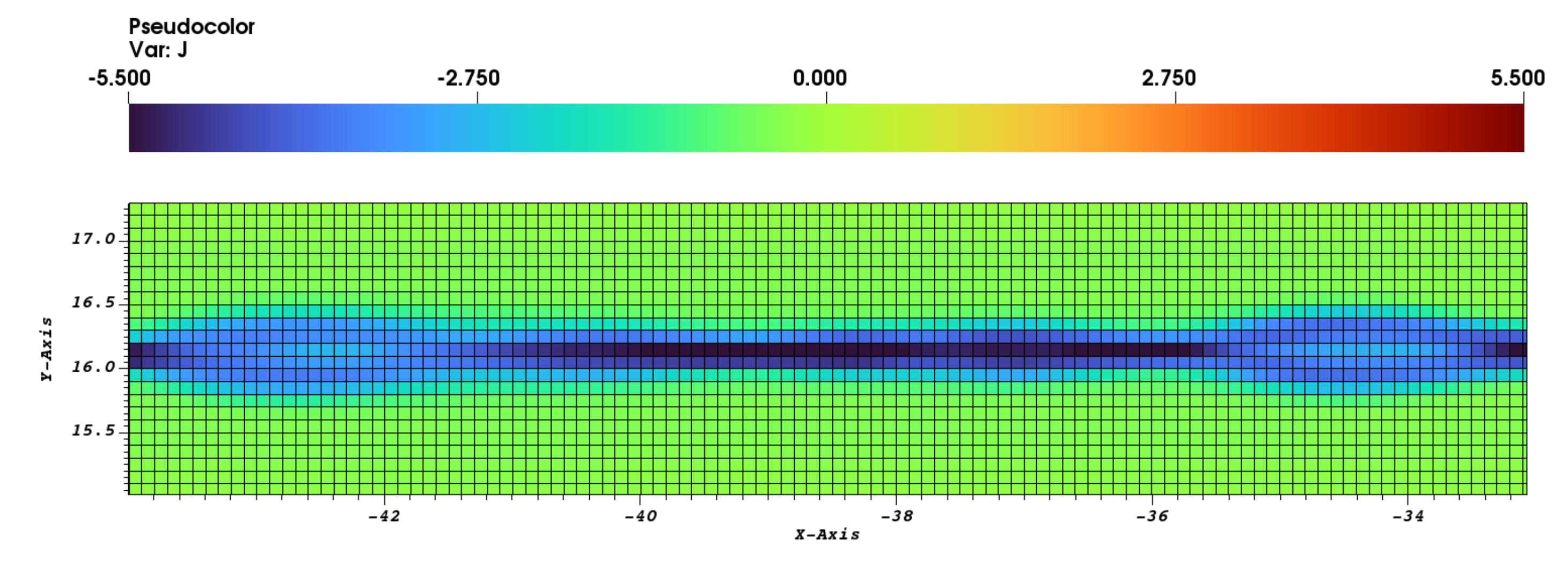}
    \caption{A close up view of the current density \textbf{J} and the grid layout in the plasmoid unstable region just at the start of the disruption process ($t/\tau_{\rm A} \sim  180$). The grid cell sizes shown in the image are is $\Delta x= \Delta y = 0.1$.}
    \label{fig:res_test_grid}
\end{figure*}

Figure \ref{fig:res_test} shows the change in magnetic energy density of the S0 domain with time. We see initially, that the lowest resolution run (marked in red) is distinct from the higher resolution runs even during the early phases of the evolution. The simulation having a resolution of $640 \times 960$, appear to converge during the early times of the evolution, i.e. $t/\tau_{\rm A} <  120$, but deviates from the higher resolution runs beyond this time. The standard resolution employed throughout our study is $1280\times1920$ which is represented by the magenta curve in Figure \ref{fig:res_test}. Two higher resolution runs were performed with a resolution 1.5x and 2x higher than this standard resolution. As seen from Figure \ref{fig:res_test}, the time of onset of the plasmoid instability is identical for the higher resolution runs. This indicates that the disruption of the current sheets at the critical thickness of $\Delta x \sim 0.37$ is indeed captured correctly and is not an artefact of resolution. We have also checked that the thickness of the current sheet at disruption remains the same even with a resolution of $2560\times 3840$ (twice the standard resolution). Also, we plot the two time durations over which the reconnection rates were measured, by the blue and green shaded strips in Figure \ref{fig:res_test}. We observe that the curve depicting the standard resolution show a very similar trend during these periods when compared to the curve corresponding to twice the standard resolution indicating that the results of the reconnection rate would stand unmodified even for a resolution of twice the standard one.

As per the quantification given in Pucci and Velli (2014) \cite{Pucci_2013},  we expect our current sheets to fragment beyond a critical width of $\Delta x \sim 0.33$ and the resolutions employed that can resolve this thickness with at least a few grid cells are expected to converge on the onset time. Thus, the resolutions of $1280\times1920$ and higher agree upon the onset time of the plasmoid instability. The evolution of the setups at higher resolutions, however, show a slight difference \textit{after} the onset of the plasmoid instability due to the inherent turbulence associated with such a fast and bursty process. Regardless, the final relaxation phase ($t/\tau_{\rm A} >  450$) is also similar for all the systems at the end of the global reconnection process.

Figure \ref{fig:res_test_grid} shows a highly zoomed in view of a portion of the plasmoid unstable current sheet just at the beginning of the disruption process. We have overlaid the image with the grid cell layout of our standard resolution. As evident from the plot, the current sheet is resolved by at least 3 to 4 grid cells during the onset of the plasmoid instability i.e when it is the thinnest. The smallest plasmoids in the domain during this time are also resolved by 5 to 7 grid cells. One should note that after the disruption, the plasmoids get larger with time and are thus being covered by a higher number of grid cells and Figure \ref{fig:res_test_grid} is in fact showing the smallest scale plasmoids of the domain. We therefore conclude that our rather modest resolution of $1280\times1920$ has sufficiently captured the dynamics of the smallest scale structures in the domain. We also assert that employing a higher resolution would indeed help in better resolving these very small scale structures but would be much more computationally intensive. Also, such an exercise would not have any significant impacts on the principal results of our study concerning the reconnection rate as well as on the mechanisms of particle acceleration.

\end{document}